# Neural-Network-based NLOS Identification of Angular Clusters at 60-GHz

Pengfei Lyu, Aziz Benlarbi-Delaï, Zhuoxiang Ren, *senior member*, *IEEE*, and Julien Sarrazin, *senior member*, *IEEE*

***Abstract*—The work in this paper identifies the nature of individual angular clusters as line-of-sight (LOS) or non-line-of-sight (NLOS) in indoor millimeter-wave channels. The proposed technique utilizes the channel knowledge that is readily available from a beam training process in directional antenna-based communications. In particular, the behavior of five different channel metrics, namely the angular covariance, the time-domain, and frequency-domain channel kurtosis, the mean excess delay, and the RMS delay spread, is analyzed using maximum likelihood ratio and artificial neural network. A noticeable difference between LOS and NLOS clusters is observed and assessed for identification. Hypothesis testing shows errors as low as 0.02-0.003 in simulations and 0.04-0.07 in measurements at 60 GHz in indoor short-range environments.**

## I. Introduction

Outdoor wireless localization such as Global Positioning System (GPS) experiences a great success due to its large-scale application. With the booming of the Internet-of-things (IoT), indoor wireless positioning applications are found in many fields such as healthcare, industrial automation, and smart environment [1]. Localization strategies based on existing wireless communication technologies take advantage of saving infrastructure costs. Thanks to their wide bandwidth, millimeter-wave (mm-wave) communications appear as a promising candidate for accurate indoor wireless localization. Operation at mm-wave is included in 5G standard. 3GPP release 15 [2] has defined the use of bandwidth in the 24.25-40 GHz range while release 17 [3] is currently considering frequencies in the 52.6-71 GHz spectrum, including the 60 GHz license-free band. The 60 GHz band is also used in IEEE 802.11ad [4] and IEEE 802.11ay [5] standards for indoor communications based on channel study [6]. In these standards, the total bandwidth at 60 GHz is 8.64 GHz [7], which can provide high positioning accuracy.

Manuscript received September 23, 2021. This work was supported in part by National Key R&D Program of China under Grant 2017YFB020350, National Science Foundation of China under Grants 61501454. This work was performed within NOVIS60 project supported by CEFIPRA (Indo-French Center for the Promotion of Advanced Research).

Pengfei Lyu is with the Institute of Microelectronics, Chinese Academy of Sciences, 100029 Beijing, China; School of Microelectronics, University of Chinese Academy of Sciences, 100049 Beijing, China; (e-mail: lvpengfei@ime.ac.cn).

Pengfei Lyu, Julien Sarrazin, Aziz Benlarbi-Delaï, and Zhuoxiang Ren are with Sorbonne Université, CNRS, Laboratoire de Génie Electrique et Electronique de Paris, 75252, Paris, France and with Université Paris-Saclay, CentraleSupélec, CNRS, Laboratoire de Génie Electrique et Electronique de Paris, 91192, Gif-sur-Yvette, France (e-mail: julien.sarrazin@sorbonne-universite.fr).

Unfortunately, the attenuation of millimeter waves by typical scatterers, such as metal and wood objects [8] and human bodies [9-11], is very high so that the direct path is frequently obstructed by outdoor random building [12, 13] or indoor human activity [14, 15]. The probability of mm-wave communications to operate over indirect paths is thus significantly enhanced, which in turn leads to biased localization metrics such as received signal strength (RSS), angle of arrival (AOA), time of arrival (TOA) [16-18], and time difference of arrival (TDOA) [19]. Triangulation-based positioning needs a direct path between transmitters (Tx) and receivers (Rx) to be accurate [16, 20]. Since multipath effect in the 3.1-to-10.6-GHz UWB band is rich [21-23], signals that have propagated along different paths sum up at Rx. In that case, two steps are implemented to mitigate the impact of indirect paths on the localization process [24]. First, the receiver identifies if a direct path exists with a given transmitter, i.e., line-of-sight (LOS) scenario, or not, i.e., non-line-of-sight (NLOS) scenario. Second, the receiver uses the signal received from LOS transmitters only to localize itself and tries to mitigate the error caused by indirect paths in the LOS channel. Signal received from NLOS transmitters are completely discarded, due to the absence of correct position information provided by the direct path. LOS and NLOS channels can be discriminated against by RSS [25, 26] or the Rician *K*-factor for instance [27]. This later can be estimated by fitting the measured data with Ricean distribution or by a verified second-order moment of the received power intensity [28, 29]. Then, NLOS identification makes a statistical decision [30] based on the difference between the parameters in LOS and NLOS channels [21, 24]. Since RSS and K-factor typically depend on the Tx-Rx range, additional features are used in the literature. For instance in [31], the mean total energy, maximum power, amplitude and rise time of the first peak, excess delay, RMS delay spread, and the received signal kurtosis (i.e., fourth-order moment) are introduced as identification metrics. While only a subset of those metrics is found to be optimal for identification in [31], the work in [32] highlights that using more metrics enables an identification more robust to various environments. Thanks to multiple-antenna systems, spatial metrics based on AOA have been used successfully to perform identification [33]. To jointly make use of several metrics in the identification process, common parametric hypothesis testing methods such as maximum likelihood ratio (MLR) [32], and non-parametric ones [31, 33-37], such as support vector machine (SVM) [31], relevance vector machine (RVM) [34], and artificial neural network (ANN) [35], are typically employed. More recently, artificial intelligence techniques have been used to perform channel characterization and identification, as summarized in this survey [38, 39]. To facilitate practical implementations, [37] uses only the channel state information (CSI) available in 5G communications to perform identification. After NLOS identification, some regression methods such as least square (LS) or maximum likelihood (ML) [40] are used to mitigate the positioning error in NLOS transmissions.

Different from the rich multipath channel at low-frequency band, the mm-wave channel is sparse [41, 42], which exhibits new NLOS identification challenges. To enhance the signal coverage affected by serious blockage, beamforming [43, 44], whether fully-digital [45] or hybrid [46], is used to look for path(s) with high signal-to-noise-ratio (SNR) according to, typically, a discrete multi-beam codebook [47]. Identified high-SNR path(s) are used to establish the communication link, which also provides higher degrees of freedom for outdoor multi-user scenarios. This beam training strategy is already included in IEEE 802.11ad and IEEE 802.11ay [7, 48, 49] and is attracting more attention in recent years to accelerate it [50, 51]. Unfortunately, while current beam training finds high-SNR links for communication, it is not meant for identifying its nature (i.e., LOS or NLOS), as required for localization. This shortcoming is addressed in this paper.

Although many studies report on indoor [52-54] and outdoor [55-58] localization based on mm-waves, only a few works [59-63] are dealing with NLOS identification at mm-wave frequencies. To maintain a low-complexity system, the work in [59, 60] uses a simple energy detector to mitigate NLOS components in TOA estimations. However, its robustness to various environments is not validated with measurements. The work in [61] assumes multiple base stations communicating with a user. However, the channel model used is obtained with a simplified ray tracing, not accounting for diffuse scattering effects, which is however a significant contribution at mm-wave [64]. [63] performs identification by exploiting several metrics thanks to a learning process based on a random forest algorithm. While the method appears to be robust to small SNR, the work is solely based on simulations. The work in [61] is to author's best knowledge the only study based on actual experiments at mm-wave. Based on 28 GHz measurements with quasi-omnidirectional antennas, it uses similar identification approaches than in UWB. Therefore, as well as all mm-wave identification, studies [59-63] do not consider the directional behavior of the channel.

As a general observation, all studies in the UWB or mm-wave band aim to identify whether a quasi-omnidirectional channel contains a LOS path. In beamforming mm-wave communications, the channel contains less information since it is spatially filtered by the directional antenna pattern, which could jeopardize identification techniques. Furthermore, after the beam training process, the antenna beam is steered towards a particular angular cluster whose nature (LOS or NLOS) needs to be known for localization purpose.

Consequently, the main contribution of this paper is the NLOS identification of *individual angular clusters*. This work leverages the multiple observations of the channel available thanks to beam training to first identify all the angular clusters of a 60 GHz indoor channel. Each of them is then classified as LOS or NLOS. Furthermore, this paper presents for the first time, to author's best knowledge, experimental identification results obtained using directional antennas at mm-wave. The reminder of the paper is organized as follows. In section II, the principle of spatial NLOS cluster identification is presented, including the different physical features between LOS and NLOS transmission, the metrics to identify, and the methods of identification. The method is implemented in simulation in section III based on the IEEE 802.11ad channel model and results are given and discussed. An experimental setup at 60 GHz is used in section IV to validate the proposed approach. Finally, section V draws conclusions and gives some perspectives of this work.

## II. Principle of Spatial NLOS Identification

The NLOS identification proposed in this paper intends to classify each cluster of a channel as LOS or NLOS. A LOS cluster is defined as an angular cluster that contains the LOS path with possibly some NLOS multipath components (MPC). An NLOS cluster is defined as an angular cluster that contains only NLOS MPC. The identification is performed by analyzing the behavior of five metrics, namely the angular covariance, the time-domain and frequency-domain kurtosis, the mean excess delay, and the RMS delay spread. This analysis is done in the elevation/azimuth angular domain using an estimated power angular spectrum (PAS). The metrics are calculated for all individual angular clusters and their statistical distributions are fitted with some distributions whose parameters allows for discrimination (LOS or NLOS). The general procedure is summarized in Fig. 1 and is detailed in the following subsections.

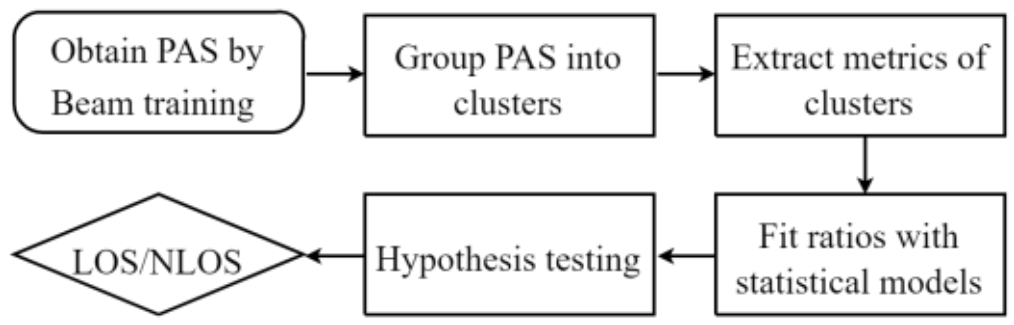

Fig. 1. Flowchart of spatial NLOS identification with beam training.

### A. Power Angular Spectrum

The beam training procedure to align Tx and Rx beams consists typically in a channel estimation over the whole angular space in both azimuth ($\phi$) and elevation ($\theta$) directions. For instance, the IEEE 802.11ad standard [7] introduces a beamforming protocol in which the Tx beam sweeps over the whole angular space, while the Rx antenna pattern is kept quasi-omnidirectional. The Rx antenna sweeps then its beam while the Tx pattern is kept quasi-omnidirectional and the link exhibiting the highest link budget may be used for communication. This beam training is performed according to a fixed codebook, in which the angular sweep is done using approximately equal intervals throughout the entire space. Beside enabling establishing a link, such procedure enables the transmitter and/or the receiver to have the knowledge of the spatial representation of the channel, which is leveraged to perform NLOS identification. Assuming the Tx antenna pattern omnidirectional, the sampled channel impulse response (CIR) estimated at Rx can be written as a function of excess delay $\tau$:

$$\hat{h}^i(\tau) = \sum_{l=1}^{L}\sum_{k=1}^{K} \alpha_k^i \operatorname{sinc}\left[2B\left(\tau_k - l\Delta\tau\right)\right]\delta\left(\tau - l\Delta\tau\right) + n^i(\tau) \quad (1)$$

where $i$ is the beam index, $L$ is the number of estimated channel taps, $K$ is the number of multipath components (MPC), $2B$ is the finite bandwidth of the receiver which sets its time resolution to $\Delta\tau = 1/2B$, $\delta(\cdot)$ is the Dirac delta function, $n(\tau)$ is an additive white Gaussian noise (AWGN), and $\alpha_k^i$ is the complex coefficient of the $k$-th MPC given by:

$$\alpha_k^i = \sqrt{g_r^i\left(\theta_k, \phi_k\right)}a_k \exp\left(j\varphi_k\right) \quad (2)$$

where $g_r^i(\theta_k, \phi_k)$ is the Rx antenna gain as experienced by the $k$-th MPC whose elevation and azimuth AOA are $\theta_k$ and $\phi_k$ respectively, and $a_k$ and $\varphi_k$ are the corresponding amplitude and the carrier phase. From (1), the receiver can calculate the channel total power, $\hat{P}$, at each beam direction $i$ defined by $(\theta^i, \phi^i)$, as:

$$\hat{P}\left(\theta^i, \phi^i\right) = \frac{1}{T}\int_0^T \left|\hat{h}^i(\tau)\right|^2 d\tau \quad (3)$$

where $T$=$L\Delta\tau$ is the estimated channel duration. Steering the Rx antenna beam leads therefore to an estimation of the PAS, defined as the angular distribution of the channel power. An example of the PAS obtained with the IEEE 802.11ad conference room scenario [4] is shown in Fig. 2 (Rx antenna beamwidth of 10° and step size of 5°). Since the 60 GHz channel is sparse [42, 65], the PAS in Fig. 2 is composed with high power clusters and a low power background. The first step is to identify and distinguish the different clusters in the PAS onto which an analysis is carried out in order to determine their nature.

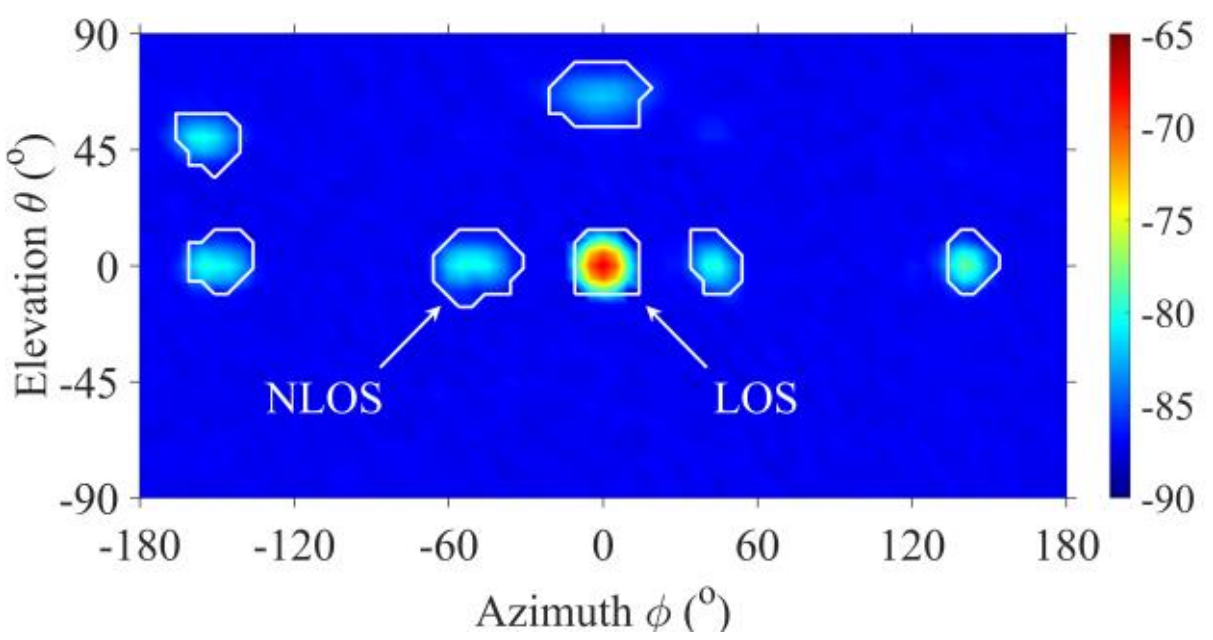

Fig. 2. Estimated power angular spectrum (in dB) obtained with beam training at Rx.

### B. PAS Clustering

In propagation channel modeling, peaks, representing impinging plane waves, are firstly extracted from the channel response and clustering is then performed on a discrete representation of the channel, whether in time or in space [66]. In this paper, Fig. 2 is directly clustered using an image processing-based method described in [67]. The PAS is processed as a grayscale image. One pixel $i$ corresponds to one Rx antenna beam direction $(\theta^i, \phi^i)$ and its value is given by (3). The pixel size is therefore equal to the steering angular step. To avoid over-segmentation, the PAS image is first filtered with a set of mathematical morphology operations, namely, opening, closing, and reconstruction [67]. The technique identifies then the illuminated foreground of the PAS 2D map, i.e., the clusters, from the lower intensity background thanks to several operations including watershed segmentation. Watershed segmentation is an unsupervised clustering method that find the local minima centers and local maxima boundaries of clusters. In addition to run fast, it does not require the a priori knowledge of the cluster number and the use of a power threshold to remove the background noise such as in [68]. Furthermore, by avoiding any high-resolution MPC extraction, [67] preserves the spatial shape of clusters which is of importance for NLOS identification. For each cluster $n$, it results in the identification of its contour as shown in Fig. 2. It is then straightforward to determine the set of pixels $i \in \mathcal{Q}_n$ that belong to a same cluster $n$.

### C. Spatial-Domain Metric: $R_p$

The LOS cluster, located in the center in Fig. 2 ($\theta = \phi = 0°$), exhibits a rotational symmetry (according to the Rx antenna beam shape). However, in NLOS clusters, this symmetry is usually lost. To illustrate this physical phenomenon, a canonical example of Rx beam sweeping simulated with the method of moments is shown in Fig. 3. In Fig. 3 (a), the received power angular distribution is not affected by scattering objects in the LOS transmission. Thus, the PAS exhibits a shape identical to the Rx radiation pattern. However, in Fig. 3 (b), the PAS is randomly deformed by the distributed scatterers. Even when reflected by smooth reflectors, NLOS clusters are affected as seen in Fig. 3 (c). So, the angular spreading experienced by NLOS clusters is likely to be different along $\phi$ and $\theta$.

The spatial distribution of $\hat{P}(\theta^i, \phi^i)$ is therefore assessed to quantify the difference between LOS and NLOS clusters. By analogy with co-variance of joint probability distribution of two variables, a similar covariance matrix is introduced to describe the shape of the 2D angular clusters while being independent of the absolute received power:

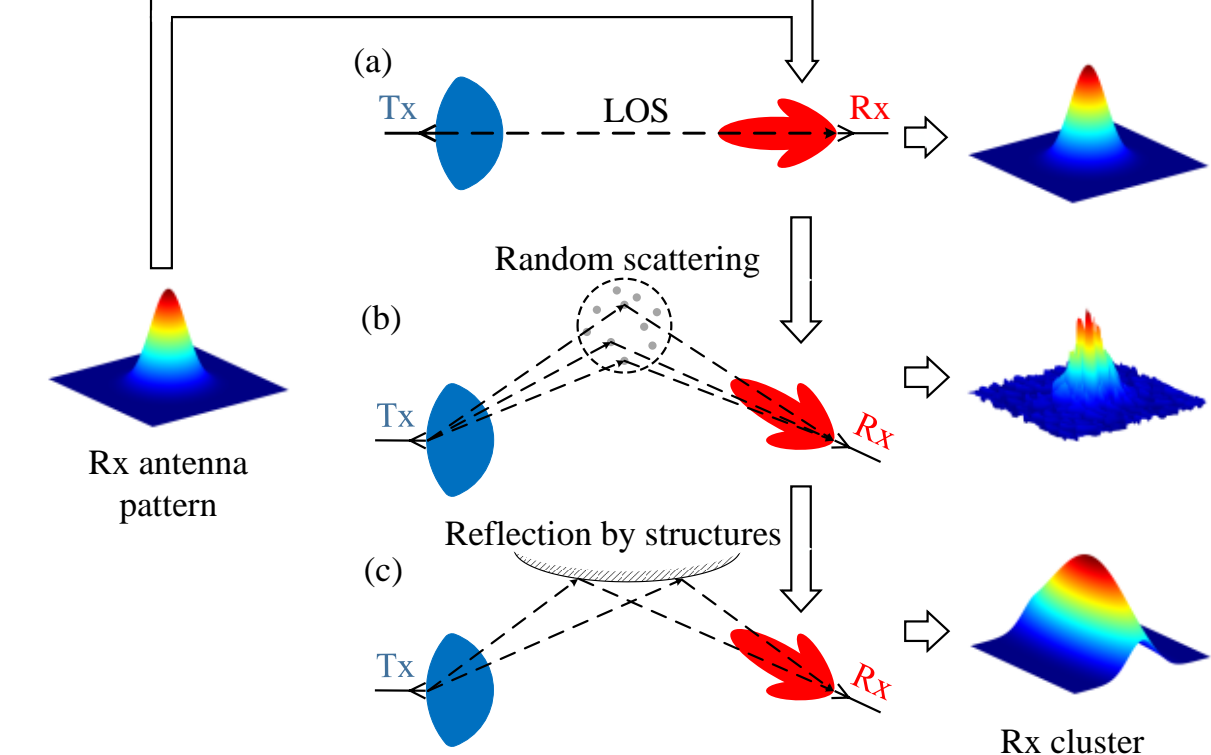

Fig. 3. Deformation of clusters due to: (a) LOS transmission; (b) random scattering; (c) mirror-like reflection.

$$\Sigma_\sigma = \begin{bmatrix} \sigma_{11}^2 & \sigma_{12}^2 \\ \sigma_{21}^2 & \sigma_{22}^2 \end{bmatrix} \tag{4}$$

with $\sigma^2$, the angular covariance weighted with power of a given cluster $n$, given by:

$$\sigma_{11}^2 = \frac{\sum \left(\phi^i - \bar{\phi}^n\right)^2 \hat{P}\left(\phi^i, \phi^i\right)}{\sum \hat{P}\left(\phi^i, \phi^i\right)} \tag{5}$$

$$\sigma_{12}^2 = \sigma_{21}^2 = \frac{\sum \left(\phi^i - \bar{\phi}^n\right)\left(\theta^i - \bar{\theta}^n\right) \hat{P}\left(\phi^i, \theta^i\right)}{\sum \hat{P}\left(\phi^i, \theta^i\right)} \tag{6}$$

$$\sigma_{22}^2 = \frac{\sum \left(\theta^i - \bar{\theta}^n\right)^2 \hat{P}\left(\theta^i, \theta^i\right)}{\sum \hat{P}\left(\theta^i, \theta^i\right)} \tag{7}$$

for $i \in \mathcal{Q}_n$. $\theta^i$ and $\phi^i$ are the azimuth and elevation angle of the $i$-th pixel. $\bar{\theta}^n$ and $\bar{\phi}^n$ are the mean angles of the $n$-th cluster. By weighting the shape of the clusters by $\hat{P}(\theta^i, \phi^i)$, the covariance matrix offers a simple mean to quantify its spatial distribution and has been found exhibiting superior identification results than higher order moments. As observed earlier, the nature of the propagation influences the symmetry of clusters. The weighted symmetry is a measurable quantity to evaluate the shape of clusters. By analogy to principal component analysis (PCA) in statistics, the ratio $R_P$ of minimum eigenvalue over maximum eigenvalue of the covariance matrix (4) can be used to characterize the spatial symmetry of the power-weighted covariance of individual clusters:

$$R_P = \frac{\min\left(\lambda_1, \lambda_2\right)}{\max\left(\lambda_1, \lambda_2\right)} \tag{8}$$

where $\lambda_1$ and $\lambda_2$ are the eigenvalues with the corresponding eigenvectors $\boldsymbol{v}_1$ and $\boldsymbol{v}_2$ through decomposition of covariance matrix (4):

$$\Sigma_\sigma = \left[\boldsymbol{v}_1 \boldsymbol{v}_2\right] \begin{bmatrix} \lambda_1 & 0 \\ 0 & \lambda_2 \end{bmatrix} \left[\boldsymbol{v}_1 \boldsymbol{v}_2\right]^{-1} \tag{9}$$

Since the variance is not a normalized moment, it increases with the Tx-Rx range. However, since it varies uniformly in any angular plane, doing the ratio in (8) makes $R_p$ range independent. One $R_p$ value is consequently obtained per cluster which is the first metric used for NLOS identification in following sections.

### *D. Time-Domain Metrics: $K_t$, $\bar{\tau}$, $\tau_{RMS}$*

Besides power, intra-cluster features in time and frequency domain also differ between LOS and NLOS clusters. An example of the power CIR clusters is shown in Fig. 4 (a)[1]. The first peak is much stronger than the other peaks in the LOS situation whereas it is not the case in NLOS where the power is more spread over time. The probability density function (PDF) of the CIR power distribution in Fig. 4 (a) is shown in Fig. 4 (b). The NLOS PDF is more spread than the LOS, which can be observed in the PDF tails.

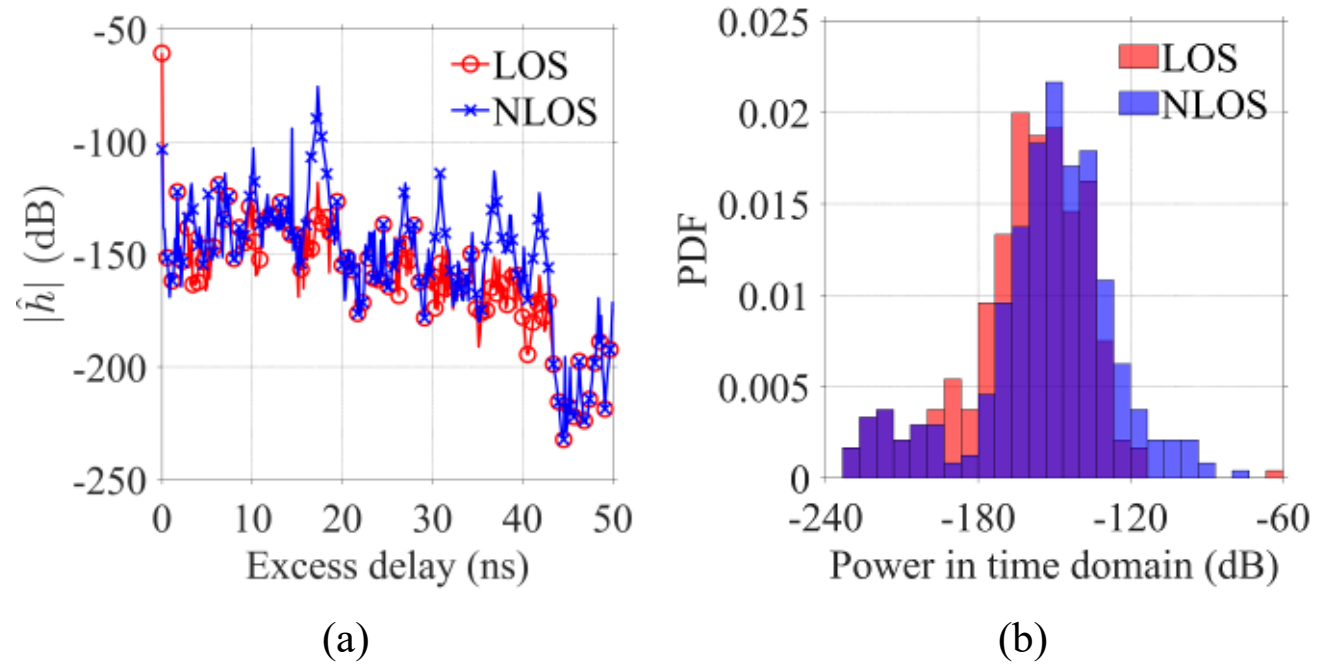

Fig. 4. (a) CIR and (b) PDF of CIR inside LOS and NLOS clusters using IEEE 802.11ad channel model of conference room scenario (noiseless case).

To obtain a metric representative of such effect and being independent to the absolute received power, the kurtosis $K_t^i$ is used to evaluate the shape of the PDF of the CIR distribution:

$$K_t^i\left(\theta^i,\phi^i\right)=\frac{E_\tau\left[\left(\left|\hat{h}^i(\tau)\right|-\mu_{\left|\hat{h}^i\right|}\right)^4\right]}{E_\tau\left[\left(\left|\hat{h}^i(\tau)\right|-\mu_{\left|\hat{h}^i\right|}\right)^2\right]^2} \tag{10}$$

where $\mu_{|\hat{h}^i|}$ is the CIR magnitude mean over excess delay $\tau$ of the $i$-th pixel and $E_\tau[\cdot]$ is the expectation operator over excess delay $\tau$. This time-domain kurtosis is therefore the second metric used in this paper to perform NLOS identification. The third and fourth metrics are the mean excess delay $\bar{\tau}^i$ and the RMS delay spread $\tau_{\text{rms}}^i$, calculated as:

$$\bar{\tau}^i\left(\theta^i,\phi^i\right)=\frac{\int_0^T \tau\left|\hat{h}^i(\tau)\right|^2 d\tau}{\int_0^T\left|\hat{h}^i(\tau)\right|^2 d\tau} \tag{11}$$

$$\tau_{rms}^i\left(\theta^i,\phi^i\right)=\sqrt{\frac{\int_0^T\left(\tau-\bar{\tau}^i\right)^2\left|\hat{h}^i(\tau)\right|^2 d\tau}{\int_0^T\left|\hat{h}^i(\tau)\right|^2 d\tau}} \tag{12}$$

Because of local random scattering experienced by multipath components while reflecting on rough surfaces, NLOS clusters are expected to exhibit larger $\bar{\tau}^i$ and $\tau_{\text{rms}}^i$ values, as typically observed in various channel environments [31].

### *E. Frequency-Domain Metric: $K_f$*

In frequency domain, the selective behavior of the channel can be observed in the channel frequency response (CFR) in Fig. 5 (a). While the CFR appears almost flat in the LOS case, it exhibits fluctuations in NLOS case. However, excluding the influence of the absolute received power, the difference is less remarkable as illustrated by the CFR PDF[2] in Fig. 5 (b). Nevertheless, LOS and NLOS PDF still exhibit slightly different shapes that are assessed in next sections using the kurtosis of the CFR as the fifth identification metric:

$$K_f^i\left(\theta^i,\phi^i\right)=\frac{E_f\left[\left(\left|\hat{H}^i(f)\right|-\mu_{\left|\hat{H}^i\right|}\right)^4\right]}{E_f\left[\left(\left|\hat{H}^i(f)\right|-\mu_{\left|\hat{H}^i\right|}\right)^2\right]^2} \tag{13}$$

where $\mu_{|\hat{H}^i|}$ is the CFR magnitude mean of the $i$-th pixel and $E_f[\cdot]$ is the expectation operator over frequency.

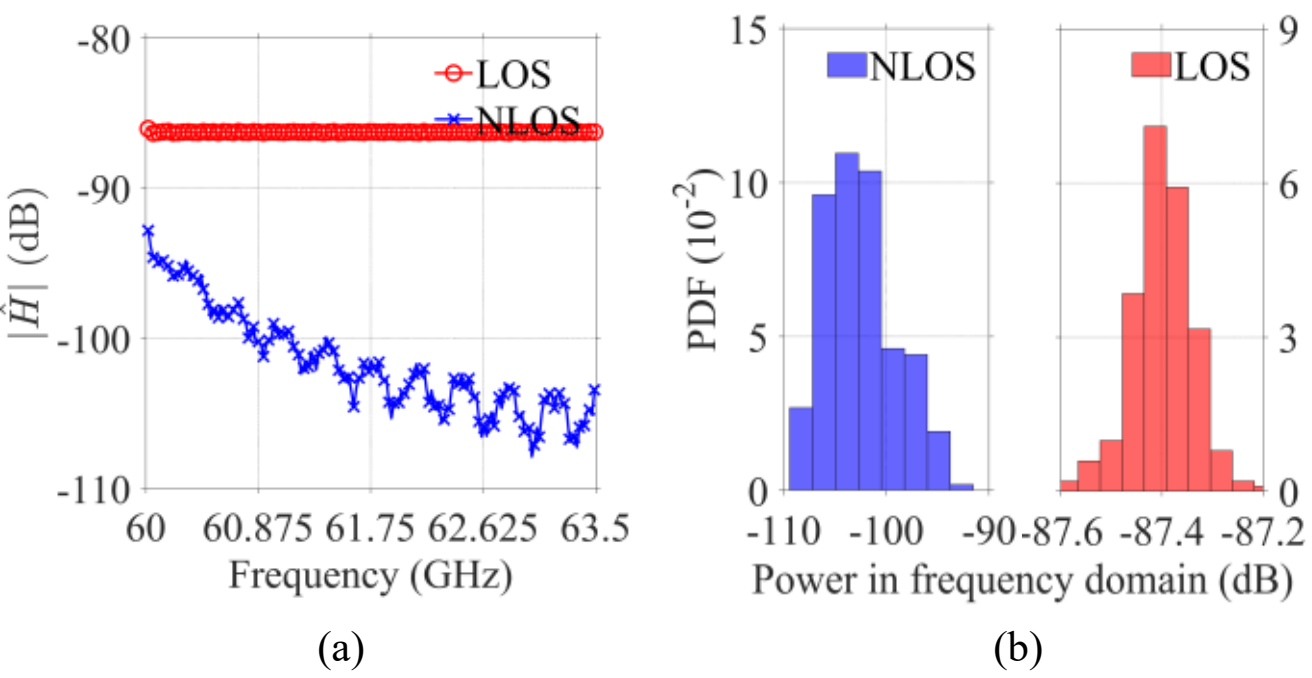

Fig. 5. (a) CFR $|H(f)|$ and (b) PDF of CFR inside LOS and NLOS using IEEE 802.11ad channel model of conference room scenario (noiseless case).

### *F. Time- and Frequency-Domain Metrics Extraction per Cluster*

While $R_p$ is evaluated per cluster, the remaining metrics are calculated for each pixel $i$. In order to obtain a single value for each identified cluster $n$, an averaging is performed such as:

$$M=\frac{1}{card\left(Q_n\right)}\sum_{i\in Q_n} M^i \tag{14}$$

where $M^i$ is either $K_t^i$, $\bar{\tau}^i$, $\tau_{\text{rms}}^i$, or $K_f^i$. Consequently, a set of five metrics $\{R_p, K_t, \bar{\tau}, \tau_{\text{rms}}, \text{and } K_f\}$ exists per cluster and is used for identification.

### *G. Statistical Model and Decisions*

Since clusters are affected by randomly distributed scatterers in an indoor scenario, the metrics associated with LOS clusters ($R_P^{los}$, $K_t^{los}$, $\bar{\tau}^{los}$, $\tau_{rms}^{los}$, and $K_f^{los}$) and NLOS clusters ($R_P^{nlos}$, $K_t^{nlos}$, $\bar{\tau}^{nlos}$, $\tau_{rms}^{nlos}$, and $K_f^{nlos}$) are stochastic variables. Therefore, the identification is a hypothesis testing: a LOS propagation hypothesis is defined as the null hypothesis, $H_0$, while the alternative hypothesis, $H_1$, is NLOS propagation:

$$\begin{aligned} &H_0\text{: LOS propagation with } R_P^{los}, K_t^{los}, \bar{\tau}^{los}, \tau_{rms}^{los}, \text{ and } K_f^{los} \\ &H_1\text{: NLOS propagation with } R_P^{nlos}, K_t^{nlos}, \bar{\tau}^{nlos}, \tau_{rms}^{nlos}, \text{ and } K_f^{nlos} \end{aligned} \tag{15}$$

Two classifiers are investigated to test the above hypothesis: MLR and ANN. The aim of MLR is to test the probability distribution from which the likelihood with the observations is sampled. The features are modeled by conditional PDFs, $f(M \mid \boldsymbol{\theta}^{los})$ and $f(M \mid \boldsymbol{\theta}^{nlos})$ with parameter vectors $\boldsymbol{\theta}^{los}$ and $\boldsymbol{\theta}^{nlos}$ for LOS and NLOS propagation, respectively.

[1] The simulation conditions are detailed in section III-A.

[2] The simulation conditions to obtain the PDF are detailed in section III-A.

The hypothesis (15) is then written as:

$$
\begin{aligned}
H_0\text{:}\ \ \hat{M}_j &\sim f\left(\hat{M}_j \mid \boldsymbol{\theta}_j^{los}\right) \\
H_1\text{:}\ \ \hat{M}_j &\sim f\left(\hat{M}_j \mid \boldsymbol{\theta}_j^{nlos}\right)
\end{aligned}
\tag{16}
$$

where $\hat{M}_j$ are the $j$-th observed metrics: $R_P$, $K_t$, $\bar{\tau}$, $\tau_{rms}$, or $K_f$. The PDFs expressions used for all $\hat{M}_j$ are the same. Specifically, generalized extreme value (GEV) are used for both simulation and measurement, as detailed in next sections. The parameters $\boldsymbol{\theta}_j$ of the above PDFs are estimated by fitting a set of training $\hat{M}_j$ with maximum likelihood estimation. With the estimated parameters, the hypothesis is tested by comparing which likelihood (LOS or NLOS) is larger at an individual testing $\hat{M}_j$ under the trained PDFs. Since the unique difference between the LOS and NLOS PDFs for a given feature $\hat{M}_j$ is the parameter $\boldsymbol{\theta}_j$, the testing becomes the comparison between the likelihood of parameters ($\boldsymbol{\theta}_j^{los}$ or $\boldsymbol{\theta}_j^{nlos}$) for a given $\hat{M}_j$. Therefore, the comparison is achieved by the likelihood ratio of the PDF parameter $\boldsymbol{\theta}_j$ at a given testing $\hat{M}_j$ ($\hat{R}_P$, $\hat{K}_t$, $\hat{\tau}$, $\hat{\tau}_{rms}$, or $\hat{K}_f$):

$$
L\left(\hat{M}_j\right)=\frac{\mathcal{L}\left(\boldsymbol{\theta}_j^{los};\hat{M}_j\right)}{\mathcal{L}\left(\boldsymbol{\theta}_j^{nlos};\hat{M}_j\right)} \underset{H_1}{\overset{H_0}{\gtrless}} 1
\tag{17}
$$

where $\mathcal{L}(\boldsymbol{\theta}_j^{los};\ \hat{M}_j)$ and $\mathcal{L}(\boldsymbol{\theta}_j^{nlos};\ \hat{M}_j)$ are likelihood functions of the statistical parameter $\boldsymbol{\theta}_j$ at a given testing $\hat{M}_j$ for LOS and NLOS propagation, respectively. Therefore, the joint likelihood ratio of $\hat{R}_P$, $\hat{K}_t$, $\hat{\tau}$, $\hat{\tau}_{rms}$, and $\hat{K}_f$ is:

$$
L\left(\hat{R}_P,\hat{K}_t,\hat{\tau},\hat{\tau}_{rms},\hat{K}_f\right)=\prod_i \frac{\mathcal{L}\left(\boldsymbol{\theta}_j^{los};\hat{M}_j\right)}{\mathcal{L}\left(\boldsymbol{\theta}_j^{nlos};\hat{M}_j\right)} \underset{H_1}{\overset{H_0}{\gtrless}} 1
\tag{18}
$$

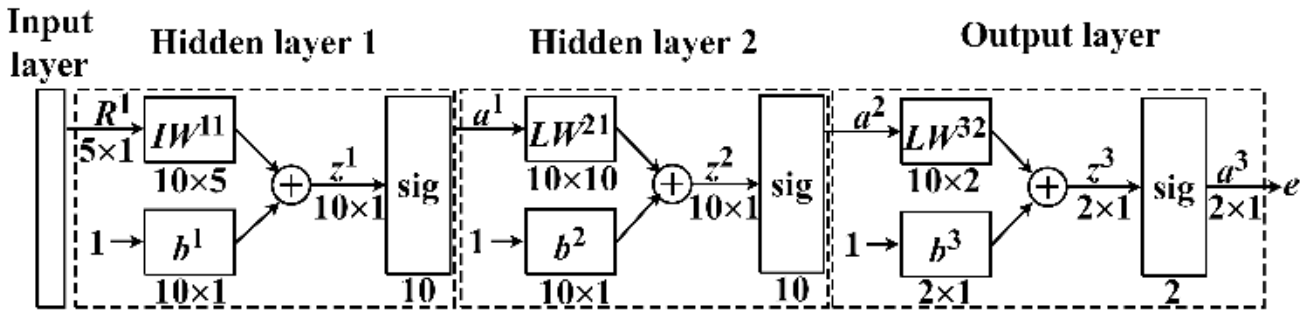

Fig. 6. Architecture of the ANN used for NLOS identification.

The second investigated method to classify the features is based on a machine learning tool, namely, the ANN. All identification metrics used in this study are calculated from a PAS directly clustered with image processing technique. Therefore, no high-resolution parameter estimation has been used. In these conditions, it was found in [33] that the ANN performs well in LOS/NLOS identification, which motivates its choice here. Its choice is further motivated by its lower complexity which makes it run faster than Random Forest (RF) in [69], Expectation Maximization for Gaussian Mixture Models (EM-GMM), LS-SVM in [70], and Convolution Neural Network (CNN) in [71]. The ANN is constructed with one input layer, two hidden layers, and one output layer, as shown in Fig. 6. It adopts a feed-forward architecture where the nodes of neighbor layers are all connected, while the nodes of nonadjacent layers are not linked (no backward cycles). Although more layers may exhibit superior classification performance, the convergence efficiency reduces accordingly while the computational complexity increases. It has been observed empirically that this 2-hidden-layer model is sufficient to exhibit interesting performance with low error probabilities as shown in Section IV.

The whole network in Fig. 6 can be expressed as:

$$
\boldsymbol{a}^3 = sig\left(\boldsymbol{LW}^{32}\cdot sig\left(\boldsymbol{LW}^{21}\cdot sig\left(\boldsymbol{IW}^{21}\boldsymbol{M}^1+\boldsymbol{b}^1\right)+\boldsymbol{b}^2\right)+\boldsymbol{b}^3\right)
\tag{19}
$$

The input is the parameter vector $\boldsymbol{M}^1 = [\hat{R}_P, \hat{K}_t, \hat{\tau}, \hat{\tau}_{rms}, \hat{K}_f]^T$. There are ten neurons in all hidden layers. The input weighting matrix in the first hidden layer $\boldsymbol{IW}^{11}$ is a 10×5 matrix. $\boldsymbol{LW}^{21}$ in the hidden layer, $\boldsymbol{LW}^{32}$ in the output layer are respectively 10×10 and 10×2 weighting matrices. $\boldsymbol{b}^1$, $\boldsymbol{b}^2$, and $\boldsymbol{b}^3$ are the bias vectors in layers 1, 2, and 3, respectively. An activation function maps the input set to an output set. For null hypothesis testing, the output set of the whole network has to be {-1, 1}, while the output of the linear weighted combination part is a real set. Therefore, the activation of output is a map $\mathbb{R} \to$ {-1, 1}. The simplest activation function is a step function. However, the step function introduces a non-derivative singularity at the origin. To avoid this issue, the activation functions in the hidden layers are both nonlinear tangent sigmoid function $sig(x) = 2 / [1 + \exp(-2x)] - 1$. The curve linearly increases in the range of [-1, 1], while approximately being constant to +1 at the top bound and to -1 at the bottom bound. The activation function of the output layer is a normalization function $sig(x) = \exp(x) / \sum\exp(x)$. It proportionally maps an input vector into a range of [0, 1]. ANN training aims to optimize the weighting matrices and the bias vectors to minimize the error $\boldsymbol{e}$ between the training result $\boldsymbol{a}^3$ and the marked label $\boldsymbol{y}$ whose values are either 0 (LOS) or 1 (NLOS) using gradient descent calculated with error-back-propagation algorithm:

$$
\boldsymbol{e} = \left(\boldsymbol{y}-\boldsymbol{a}^3\right)^2
\tag{20}
$$

## III. Simulation Results

### A. Simulation Conditions

The simulation follows the procedure in Fig. 1. The PAS is firstly obtained by beam training using the IEEE 802.11ad channel. The channel is built-in conference scenario where a pair of Tx and Rx are placed on a table at the center of a conference room with an area of 4.3 m × 3 m × 3 m [4]. The scenario is as per as the beam training strategy: the Tx antenna is omnidirectional, while the directional Rx antenna rotates across the whole angular space with a scanning step of 5°. The beam pattern of the antenna model is a single main lobe with a symmetric Gaussian shape. The half-power beamwidth (HPBW) of the Rx antenna is taken as 10°. A sampling frequency of 7 GHz ($\Delta\tau = 0.14$ ns) is used in the IEEE 802.11ad channel model generation. 2000 Monte Carlo simulations are performed to obtain enough data for a statistical description of the performance with Tx-Rx distance ranging from 2 to 3 meters. Half of the simulated channels contains a LOS cluster while the other half contains only NLOS contributions. Among the realizations, 1000 are chosen randomly for training and the other 1000 are used for hypothesis testing. The obtained PAS are clustered with the watershed algorithm in [67]. The cluster metrics are calculated according to (4) - (14). The AWGN noise in (1) is added to the generated channel with a constant power over the whole PAS. Unless stated otherwise, the SNR is set to 20 dB with respect to the largest received power among all $(\theta^i, \phi^i)$ beam positions.

### B. Statistical Characteristics of Metrics

The statistical behavior of covariance eigenvalue ratio $R_P$, time-domain kurtosis $K_t$, mean excess delay $\bar{\tau}$, RMS delay spread $\tau_{rms}$, or frequency-domain kurtosis $K_f$ is described with PDF that is fitted with GEV distributions. The expression of GEV PDF and CDF are given by:

$$f_j\left(x\,|\,\gamma_j,\mu_j,\sigma_j\right)= \begin{cases} \dfrac{1}{\sigma_j}\exp\left[-\left(1+\gamma_j\dfrac{x-\mu_j}{\sigma_j}\right)^{-1/\gamma_j}\right]\left(1+\gamma_j\dfrac{x-\mu_j}{\sigma_j}\right)^{-(1+1/\gamma_j)}, & \gamma_j\neq 0 \\ \dfrac{1}{\sigma_j}\exp\left[-\exp\left(-\dfrac{x-\mu_j}{\sigma_j}\right)-\dfrac{x-\mu_j}{\sigma_j}\right], & \gamma_j=0 \end{cases} \tag{21}$$

$$F_j\left(x\,|\,\gamma_j,\mu_j,\sigma_j\right)= \begin{cases} \exp\left[-\left(1+\gamma_j\dfrac{x-\mu_j}{\sigma_j}\right)^{-1/\gamma_j}\right], & \gamma_j\neq 0 \\ \exp\left[-\exp\left(-\dfrac{x-\mu_j}{\sigma_j}\right)\right], & \gamma_j=0 \end{cases} \tag{22}$$

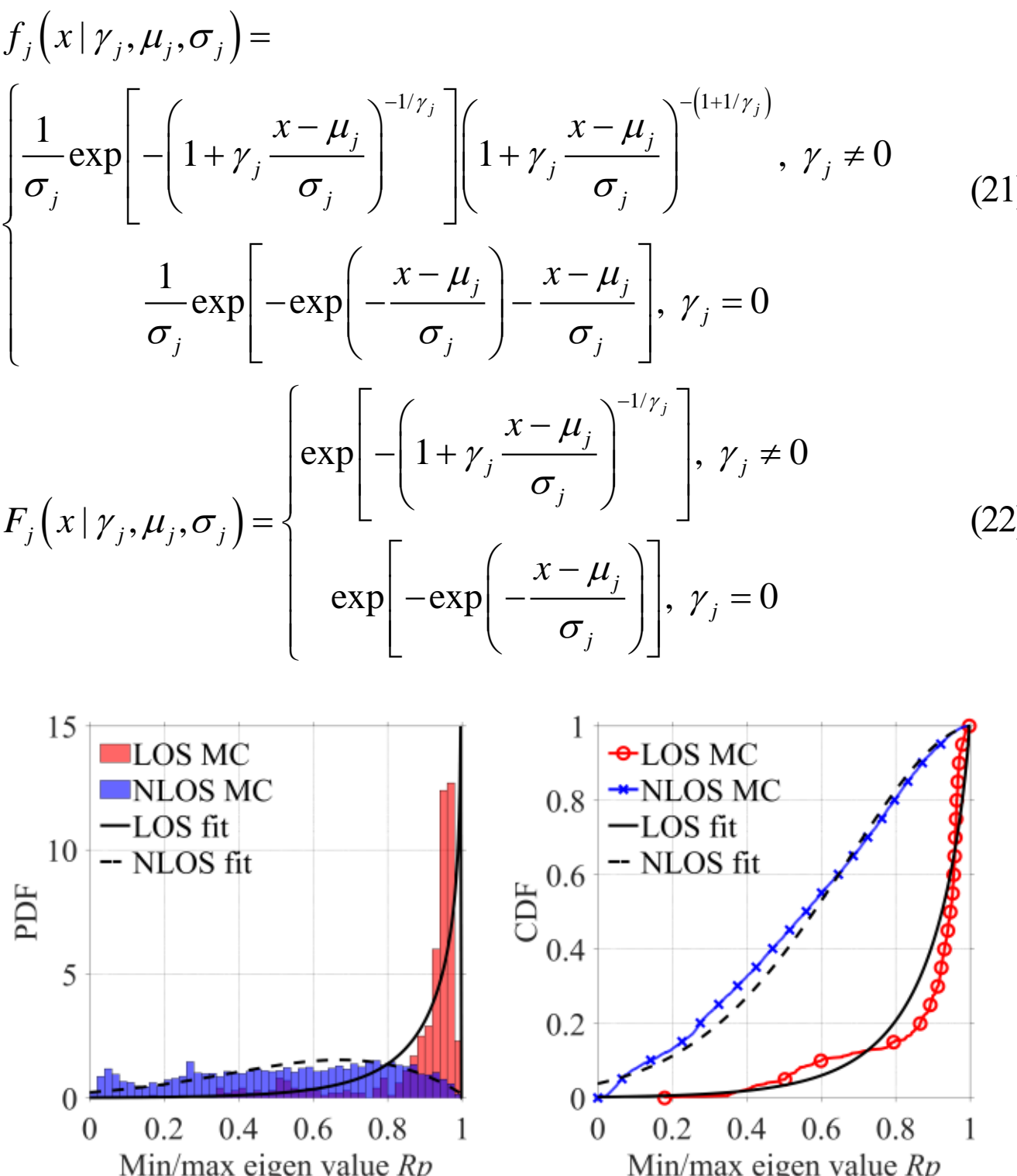

Fig. 7. PDF and CDF of simulated covariance eigenvalue ratios $R_P$.

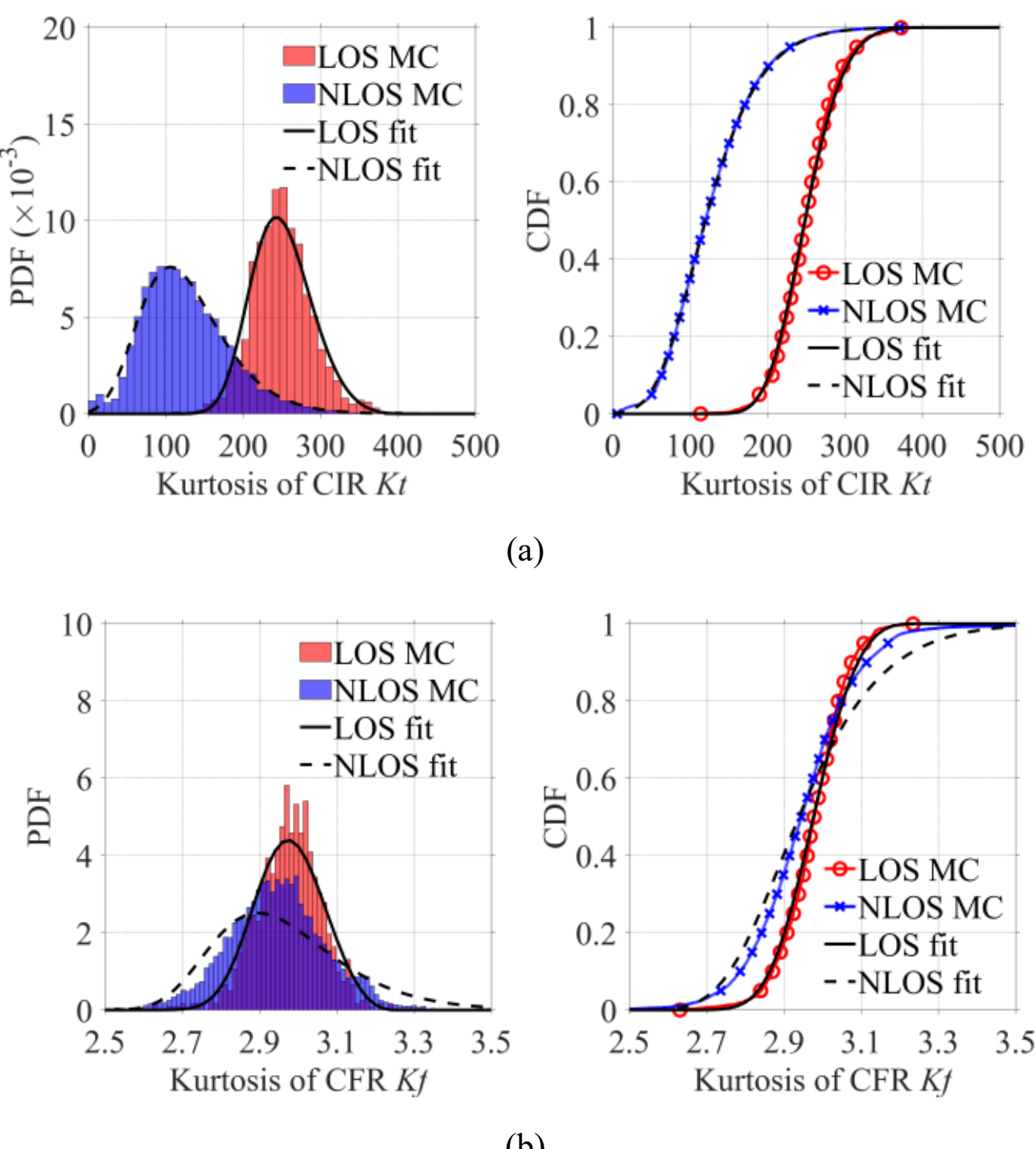

Fig. 8. PDF and CDF of simulated (a) time-domain kurtosis $K_t$ and (b) frequency-domain kurtosis $K_f$.

where $\boldsymbol{\theta_j} = \{\gamma_j, \mu_j, \sigma_j\}$ are the parameters corresponding to the $j$-th metric used in the likelihood functions in (17) and (18). GEV is used to obtain reference parameters to be used for classification in the next section. Monte Carlo simulations and fitted GEV distributions are given in Fig. 7-9 where a fair agreement can be observed. For the spatial eigenvalue ratio of power covariance matrix $\hat{R}_P$, it can be qualitatively observed in Fig. 7 that the LOS clusters exhibit a large symmetry as the PDFs tend to rapidly grow when the min/max eigenvalue ratios reach 1. In particular, it is observed that 90 % of the min/max eigenvalue ratios in LOS transmission is concentrated in a range from 0.6 to 1. For NLOS clusters, PDFs appear more uniformly distributed and there is no specific tendency for min/max eigenvalue ratios to be equal to a particular value. This shows that the spatial shape of NLOS cluster metrics is less consistent. The LOS cluster behavior is also observed in the LOS CDF curves, where an exponential-like increase occurs as ratios get close to 1. The wide min/max eigenvalue ratios spread observed in the NLOS PDF is represented by the almost linear behavior of the NLOS CDF curves. These results mean that, as expected, statistically, the channel metric $\hat{R}_P$ behaves similarly in both azimuth and elevation planes in LOS clusters, while it is stochastically affected by the random scattering in NLOS clusters and therefore behave differently in both angular planes. The LOS and NLOS CDF curves are clearly different from each other, and there is only 31% probability that LOS and NLOS PDF overlap which indicates a promising indicator for LOS/NLOS discrimination. The PDF and CDF of time- and frequency-domain kurtosis, $K_t$ and $K_f$, are shown in Fig. 8. 90% of $K_t$ values in NLOS clusters are lower than 201 as shown in Fig. 8 (a), while it is only 8.4% in LOS clusters. However, LOS and NLOS PDFs exhibit a 18% overlap. So a certain ambiguity can arise for identification, especially in the [150, 300] range. In frequency-domain, 99% of $K_f$ values in NLOS clusters are distributed in the [2.57, 3.49] range, which entirely covers the [2.68, 3.18] range that contains 99% of LOS $K_f$ values as seen in Fig. 8 (b). The LOS and NLOS CDF are therefore not too different. The clearer distinction offered by $K_t$ over $K_f$ is consistent with the example of PDFs shown in Fig. 4 (b) Fig. 5 (b).

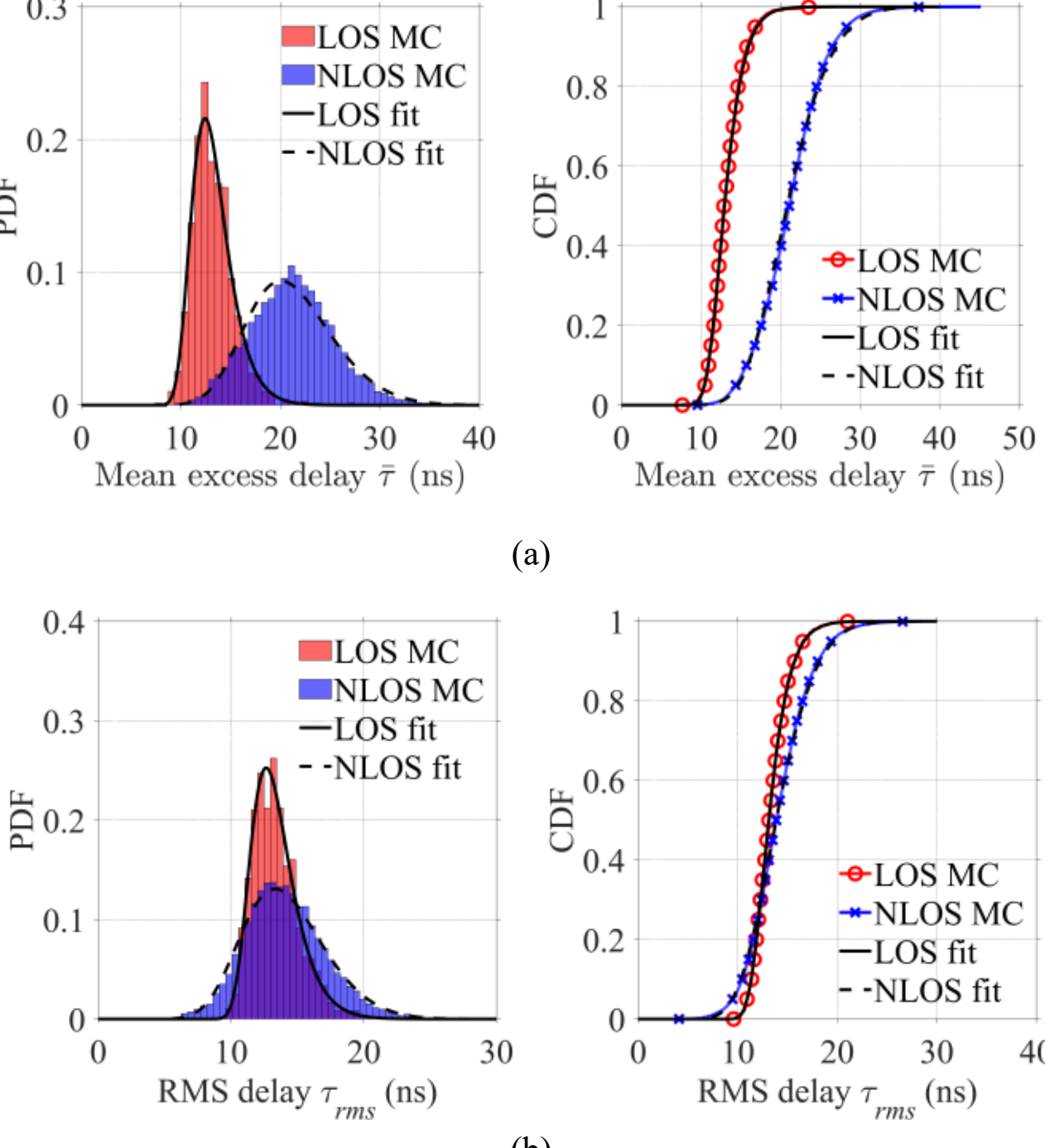

Fig. 9. PDF and CDF of simulated (a) mean excess delay $\bar{\tau}$ and (b) RMS delay spread $\tau_{rms}$.

The PDF and CDF of mean excess delay $\bar{\tau}$ and RMS delay spread $\tau_{rms}$ are shown in Fig. 9 (a), where the difference in LOS and NLOS statistical behavior can be observed. 83% of LOS clusters exhibit mean excess delays less than 15 ns whereas only 7% of LOS clusters do so. However, the difference in LOS/NLOS behavior in terms of RMS delay is not so clear as seen in Fig. 9 (b). PDFs overlap over a large area in the range 11 ns to 20 ns. So, on one hand, the diffusion in NLOS clusters introduces a larger excess delay, while on the other hand, the strong mm-wave attenuation faced by the diffused components does not increase much the RMS delay spread. The parameters of the GEV distribution and the root mean squared error (RMSE) between fitted and empirical CDF are listed in TABLE I. The RMSE of LOS and NLOS CDF is lower than 0.0841 and 0.0451 respectively.

TABLE I
PARAMETERS OF GEV DISTRIBUTION OBTAINED TO FIT THE CDF

| Metrics | Parameters | LOS | NLOS |
|---|---|---|---|
| $R_P$ | $\gamma_p$ | -1.2056 | -0.5521 |
| | $\mu_p$ | 0.8828 | 0.4835 |
| | $\sigma_p$ | 0.1381 | 0.2905 |
| | RMSE | 0.0841 | 0.0263 |
| $K_t$ | $\gamma_t$ | -0.1926 | -0.0737 |
| | $\mu_t$ | 235.3 | 102.4 |
| | $\sigma_t$ | 26.94 | 48.56 |
| | RMSE | 0.0156 | 0.0037 |
| $K_f$ | $\gamma_f$ | -0.2859 | -0.0698 |
| | $\mu_f$ | 2.946 | 2.886 |
| | $\sigma_f$ | 0.0875 | 0.1469 |
| | RMSE | 0.0204 | 0.0451 |
| $\bar{\tau}$ | $\gamma_\tau$ | -0.0460 | -0.1508 |
| | $\mu_\tau$ | 12.34 | 19.42 |
| | $\sigma_\tau$ | 1.705 | 3.983 |
| | RMSE | 0.0078 | 0.0139 |
| $\tau_{rms}$ | $\gamma_{rms}$ | -0.0414 | -0.1549 |
| | $\mu_{rms}$ | 12.58 | 12.85 |
| | $\sigma_{rms}$ | 1.457 | 2.848 |
| | RMSE | 0.0049 | 0.0079 |

### C. *Performance of NLOS Identification*

The performance of the classifiers is evaluated with the probability of incorrect decisions for a set of testing data. According to statistical decision theory, for a null hypothesis testing, the error can be classified as two types: type I error (reject the true null hypothesis $H_0$) and type II error (non-reject the false alternative hypothesis $H_1$). So according to (15), a type I error is a LOS transmission identified as NLOS and a type II error is an NLOS transmission classified as LOS. The error is evaluated by comparing the type of transmission decided by the classifiers with the actual type of transmission and results are shown in TABLE II. The performance is given for SNR = 20 dB and three angular step sizes, namely, 1°, 5°, and 10°, for which the corresponding beamwidths are respectively 5°, 10°, and 20°.

TABLE II
SIMULATED NLOS IDENTIFICATION FOR VARIOUS ANGULAR STEP SIZES (SNR = 20 DB)

| Method | $\hat{M}_j$ | 1° step | | 5° step | | 10° step | |
|---|---|---|---|---|---|---|---|
| | | I | II | I | II | I | II |
| MLR | $R_P$ | 0.050 | 0.138 | 0.154 | 0.197 | 0.200 | 0.278 |
| | $K_t$ | 0.016 | 0.060 | 0.040 | 0.125 | 0.048 | 0.173 |
| | $K_f$ | 0.080 | 0.402 | 0.152 | 0.671 | 0.100 | 0.616 |
| | $\bar{\tau}$ | 0.022 | 0.037 | 0.094 | 0.097 | 0.144 | 0.152 |
| | $\tau_{rms}$ | 0.178 | 0.237 | 0.242 | 0.447 | 0.146 | 0.477 |
| | *all* | 0.018 | 0.012 | 0.048 | 0.024 | 0.052 | 0.051 |
| ANN | *all* | 0.010 | 0.001 | 0.027 | 0.005 | 0.036 | 0.009 |

Error results differ with metrics and classification methods. Using metrics individually with MLR, the time-domain kurtosis leads to the lowest errors except for type II error with 5° and 10° steps where $\bar{\tau}$ performs slightly better. Using all metrics improves results, especially for type II errors. Interestingly, while individual metrics exhibit lower type I errors than type II, using all metrics leads to the opposite, i.e., lower type II errors than type I. ANN performs better than MLR for both type I and II errors. Indeed, different from simply comparing the probability with MLR, the relation among the metrics is discovered by ANN and gives more information regarding LOS and NLOS differences. Like with MLR, ANN exhibits lower type II errors than type I. As a general trend, the larger the angular step size, the worst the performance. However, using all metrics, identification errors remain low, less than 0.052 with MLR and 0.036 with ANN.

TABLE III shows results for lower SNR values, i.e., 15 dB and 10 dB, with a 5° step size. It can be observed that the SNR influence depends on the individual metric and the type of error. For instance, decreasing SNR increases $R_P$, $Kt$ and $\bar{\tau}$ type II errors while $Kt$ type I error remains constant. $K_f$ type II error remains very large regardless the SNR. While $\bar{\tau}$ and $\tau_{rms}$ type II errors depends on SNR, no general trend is observed. Using all metrics, identification errors remain low, with ANN performing better than MLR for all SNR values. TABLE III also reports on mixed SNR, which represents a more practical scenario. In this case, the SNR is randomly chosen between 10 and 20 dB for each channel realization for both training and testing data sets. Results show low errors which suggest robustness to SNR variation.

TABLE III
SIMULATED NLOS IDENTIFICATION FOR VARIOUS SNR (STEP SIZE = 5°)

| Method | $\hat{M}_j$ | SNR = 15 dB | | SNR = 10 dB | | Mixed SNR | |
|---|---|---|---|---|---|---|---|
| | | I | II | I | II | I | II |
| MLR | $R_P$ | 0.142 | 0.262 | 0.130 | 0.366 | 0.132 | 0.280 |
| | $K_t$ | 0.040 | 0.133 | 0.040 | 0.143 | 0.036 | 0.129 |
| | $K_f$ | 0.184 | 0.604 | 0.218 | 0.696 | 0.154 | 0.619 |
| | $\bar{\tau}$ | 0.156 | 0.144 | 0.178 | 0.130 | 0.152 | 0.142 |
| | $\tau_{rms}$ | 0.222 | 0.611 | 0.196 | 0.563 | 0.212 | 0.669 |
| | *all* | 0.064 | 0.039 | 0.032 | 0.046 | 0.050 | 0.032 |
| ANN | *all* | 0.014 | 0.003 | 0.012 | 0.009 | 0.025 | 0.003 |

TABLE II and III results are compared with those obtained in the literature from simulated quasi-omnidirectional channels. In the UWB band, it is observed that identification in indoor environments with CIR kurtosis exhibits lower error probability (0.01 in [32] to 0.19 in [33] for type I errors and 0.033 in [32] for type II errors) than with mean excess delay (0.115 and 0.137 in [32] for type I and II errors respectively) or with RMS delay spread (0.264 in [32] to 0.31 in [33] for type I errors and 0.11 in [32] to 0.276 in [35] for type II errors). This trend is similar to results shown in this paper, with $K_t$ leading to better performance than mean excess delay and RMS delay for any type of errors. Interestingly, individual cluster identification, as presented in this paper, using mean excess delay can achieve lower error probability when narrow-beam antennas are considered. In [61], identification performed with mean excess delay at 60 GHz on the omnidirectional IEEE 802.11ad channel model (with SNR = 10 dB) using random forest method achieves probabilities of 0.44 and 0.2 for type I and II errors, respectively, while identification with RMS delay spread achieves probabilities of 0.21 and 0.3 for type I and II errors, respectively. By comparing these results with TABLE III with MLR, it is observed that for the same channel model and the same SNR = 10 dB, cluster-based identification leads to lower type I error probability with mean excess delay but higher type II error probability with RMS delay. So, depending on the considered scenarios and the frequency band, the metrics that lead to the highest identification performance are not necessarily the same between identifications based on omni-directional channels or angular clusters. Overall, results obtained using several metrics with MLR are in the same order as the ones obtained by simulation on omnidirectional channels in the 3.1-6.3 GHz in [32] where 0.021 type I and 0.041 type II errors are reported.

To illustrate why ANN outperforms MLR, the correlations, $\rho^{los}$ and $\rho^{nlos}$, among the different metrics are calculated for LOS and NLOS clusters respectively. Results are shown in matrix (24) and (25). The $\rho_{ij}$ coefficient corresponds to the correlation between the $i$-th and the $j$-th metrics of the metric vector $\hat{\boldsymbol{M}} = [\hat{R}_P, \hat{K}_t, \hat{\tau}, \hat{\tau}_{rms}, \hat{K}_f]^T$.

$$\rho_{ij} = \frac{\operatorname{cov}(M_i, M_j)}{\sigma_i \sigma_j} \tag{23}$$

$$\rho^{los} = \begin{bmatrix} 1 & 0.3564 & -0.4601 & -0.0192 & 0.1680 \\ 0.3564 & 1 & -0.5108 & 0.1440 & 0.3486 \\ -0.4601 & -0.5108 & 1 & 0.5496 & -0.5421 \\ -0.0192 & 0.1440 & 0.5496 & 1 & -0.2878 \\ 0.1680 & 0.3486 & -0.5421 & -0.2878 & 1 \end{bmatrix} \tag{24}$$

$$\rho^{nlos} = \begin{bmatrix} 1 & 0.0799 & -0.0101 & -0.0825 & 0.0521 \\ 0.0799 & 1 & -0.3103 & -0.0367 & 0.0089 \\ 0.0101 & -0.3103 & 1 & 0.3371 & -0.0301 \\ 0.0825 & -0.0367 & 0.3371 & 1 & -0.2155 \\ 0.0521 & 0.0089 & -0.0301 & -0.2155 & 1 \end{bmatrix} \tag{25}$$

Some correlation values are relatively large, such as $\rho_{32}^{los}$ = -0.5108 and $\rho_{43}^{los}$ = 0.5496, meaning that some metrics are correlated between each other's. This correlation is not taken into account by the MLR approach since the probability density functions assume the metrics to be independent. However, the correlation among metrics is indeed modeled in the ANN, which better describes the relationship between the metrics $\hat{\boldsymbol{M}}$ and the nature of the cluster, thereby explaining why ANN outperforms MLR. Overall, correlation values are larger in LOS clusters than in NLOS ones, the reason being that the NLOS scattering is much more random than the LOS one.

It is to be noted that although learning can be done off-line, identification needs to be performed in real-time for localization applications. As an example, clustering a 5°-step PAS, calculating the five metrics, and identifying the clusters as LOS or NLOS takes 3.2 ms for clustering, 0.02 ms for MLR-based identification and 0.15 ms for ANN-based identification with a laptop with 2.6 GHz CPU frequency and 8 GB RAM. This appears fast enough to consider performing identification in real time.

### *D. Influence of The Training Data Set*

To investigate the influence of the training data set upon the identification performance, its size is varied according to a ratio ranging from 0.01 to 1 with respect to the previously used data set of 1000 channel realizations. A ratio of 1 corresponds to the previous case, i.e., 1000 realizations for training 1000 for testing, while a ratio of 0.01 corresponds to 10 realizations used for training and 1990 for testing. For each ratio, 100 Monte Carlo simulations are performed wherein the training samples are randomly selected. The results are shown in Fig. 10. Identification performance remains constant with ratio as low as 0.1, which suggests that 100 realizations appear sufficient for a robust training. As this ratio get lower, the MLR technique has an increasing type I error, i.e., most clusters are identified as NLOS, regardless their true nature, while ANN has an increasing type II error, i.e., most clusters are identified as LOS.

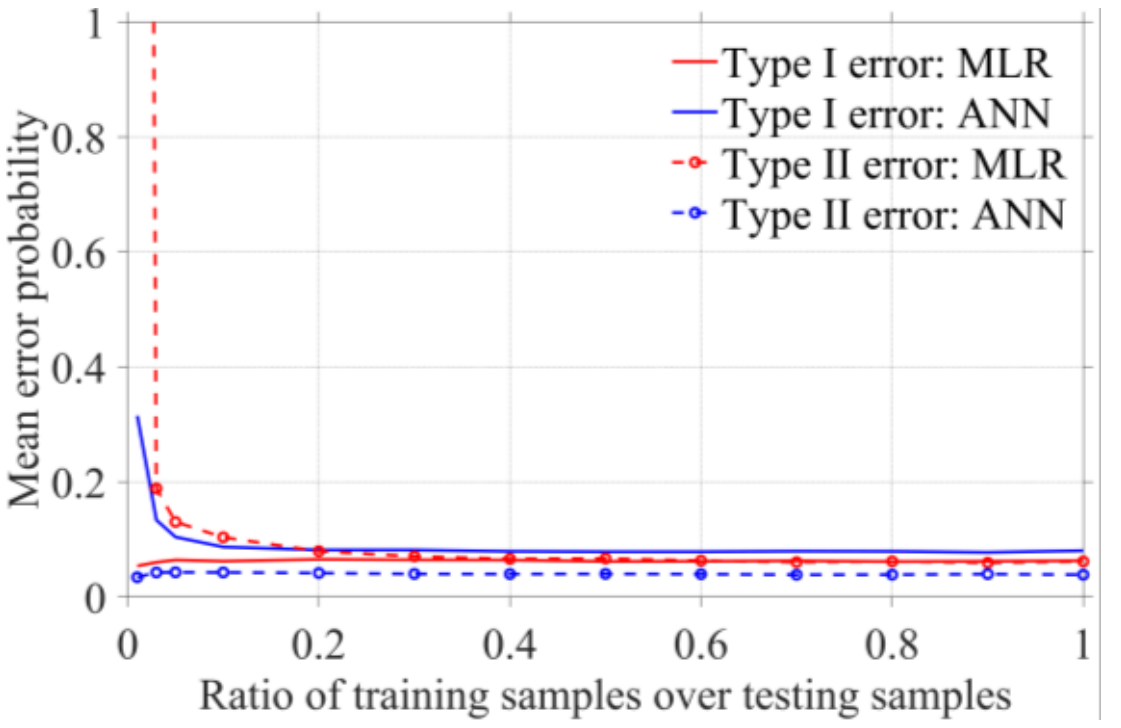

Fig. 10. Influence of the training data set size on type I and II error probabilities.

## IV. Experimental Validation

### *A. Measurement Conditions*

To validate the proposed approach, experiments are conducted in a laboratory environment. Measurements are performed in a quasi-rectangle room in a microwave wireless facilities lab at Sorbonne University. The size of the room is approximately 10.25 m × 7.52 m. The distance between the ground and the ceiling is 2.93 m. The floor plan of the measuring environment is illustrated in Fig. 11. Measurements are randomly implemented in the zones which are marked as closed circles in the floor plan: direct paths are blocked in green circles and unblocked in red circles. Both Tx and Rx are in the same zone for a given set of experiments with the distance between Tx and Rx ranging from 0.5 to 2.5 meters. A total of 100 PAS samples are measured. Since this represents a relatively small data set (a single PAS measurement about 3.5 hours), bootstrapping method is used to artificially augment it [72]. 30 samples are randomly extracted from the population as a set, while 20 samples are randomly extracted from the rest to construct the testing set. After repeating the training and testing 10 times, the final model and testing performance are obtained by averaging results over the 10 training and testing.

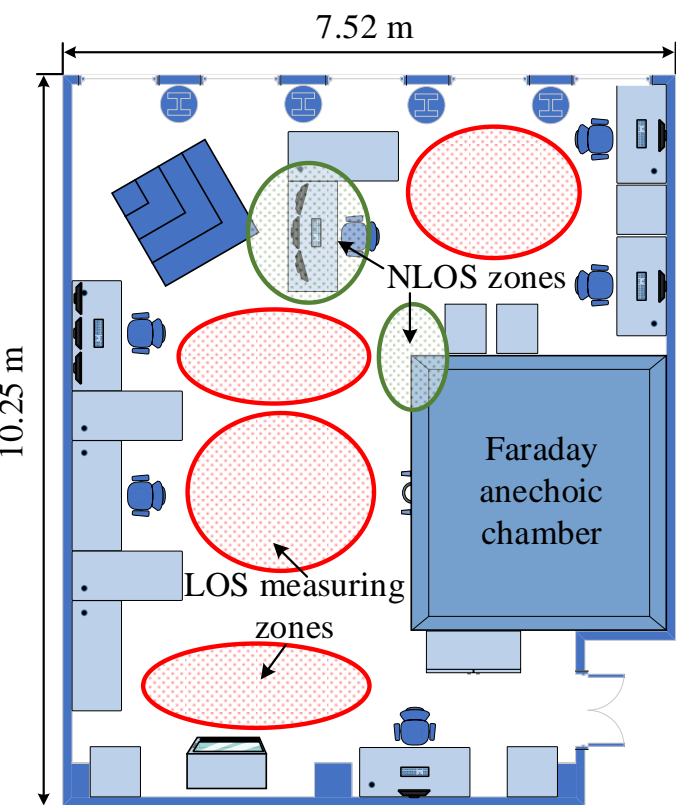

Fig. 11. Floor plan of the experiments.

The measurement configuration set-up aims at emulating a beam training strategy as shown in Fig. 12. A vector network analyzer (VNA) is used to measure the propagation channel, between Tx and Rx antennas. Tx antenna is a quasi-omnidirectional dipole antenna with 2 dB gain, while the Rx antenna is a directional horn antenna with about 10° HPBW and 24 dB gain. The beam training is achieved with Rx spatial scanning in vertical and horizonal directions by an azimuth motor and an elevation motor respectively. The sweeping steps are 5° in both directions. The parameters are listed in TABLE IV. An example of measured PAS is shown in Fig. 13, along with the identified clusters, within which the type of propagation is tested.

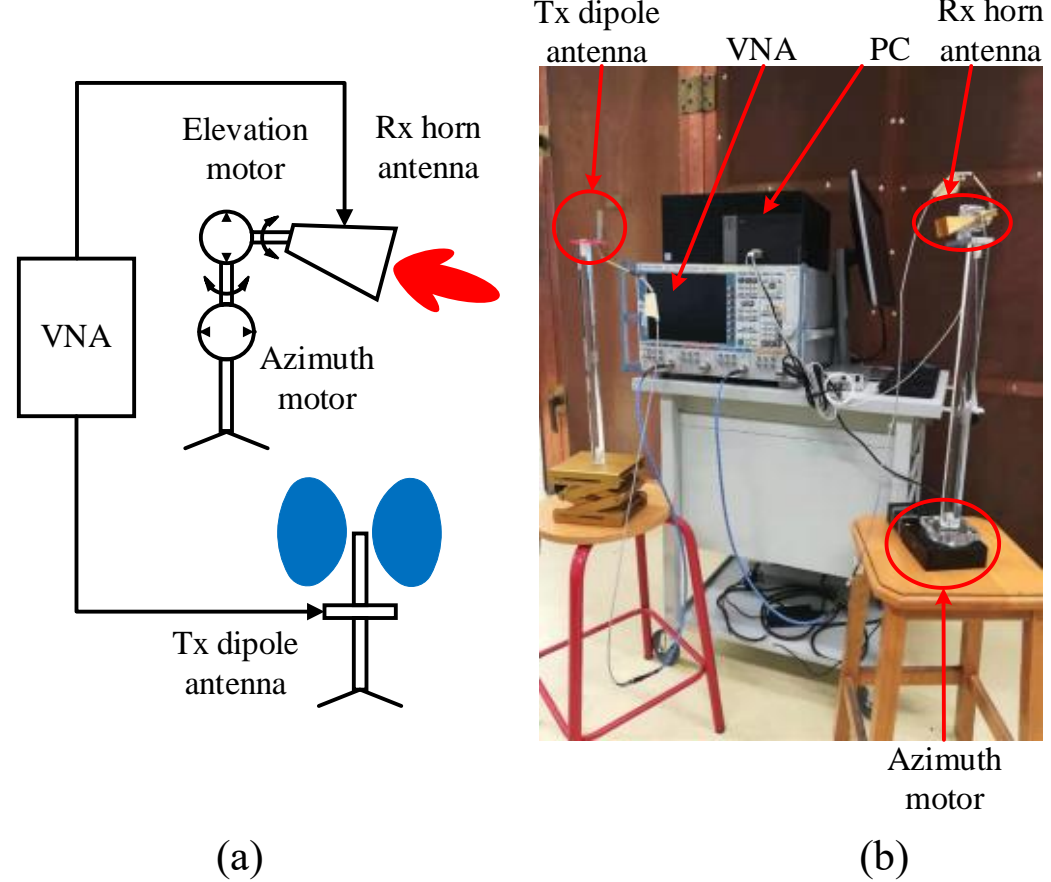

Fig. 12. (a) Schematic and (b) photo of the measurement system.

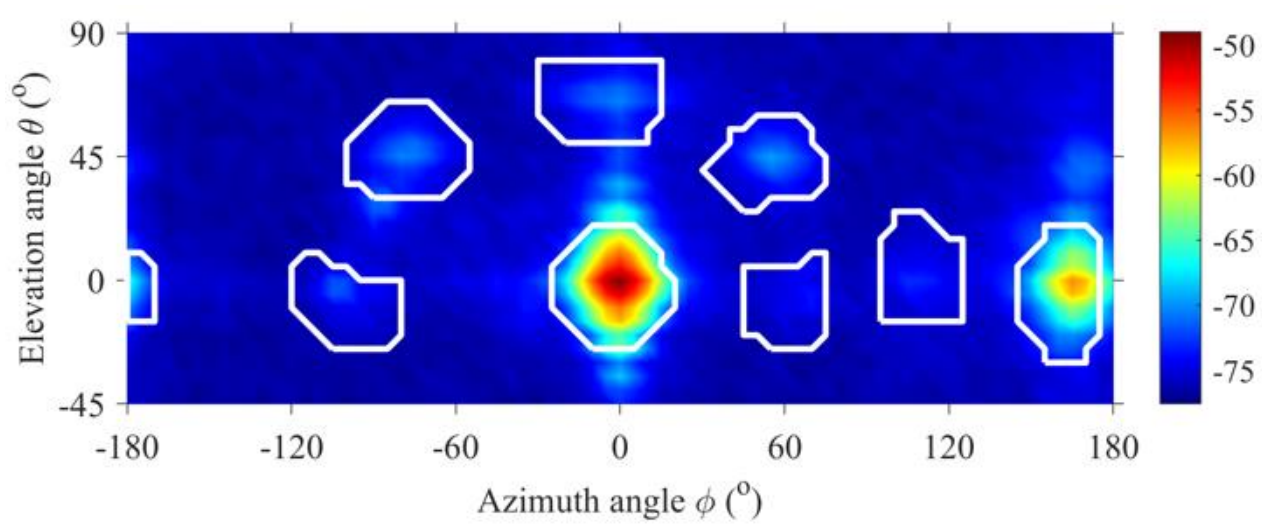

Fig. 13. Example of measured PAS in dB with clusters identified by watershed segmentation.

TABLE IV
PARAMETERS OF THE PURPOSED MEASUREMENT SYSTEM

| | |
|---|---|
| Bandwidth | 8.64 GHz |
| Time resolution Δτ | 0.12 ns |
| Frequency resolution | 11.5 MHz |
| Transmit power | 4 dBm |
| Noise level | -100 dBm |
| Rx beam width (E/H plane) | 10.1°/ 13.1° |
| Tx beam width (E/H plane) | 360 °/ 60 ° |
| Tx antenna gain | 2 dB |
| Rx antenna gain | 24 dB |
| Sampling range in azimuth | [-180°, 180 °] |
| Sampling range in elevation | [-45 °, 90 °] |
| Spatial sampling interval | 5° |

## *B. Statistical Characteristics of Metrics*

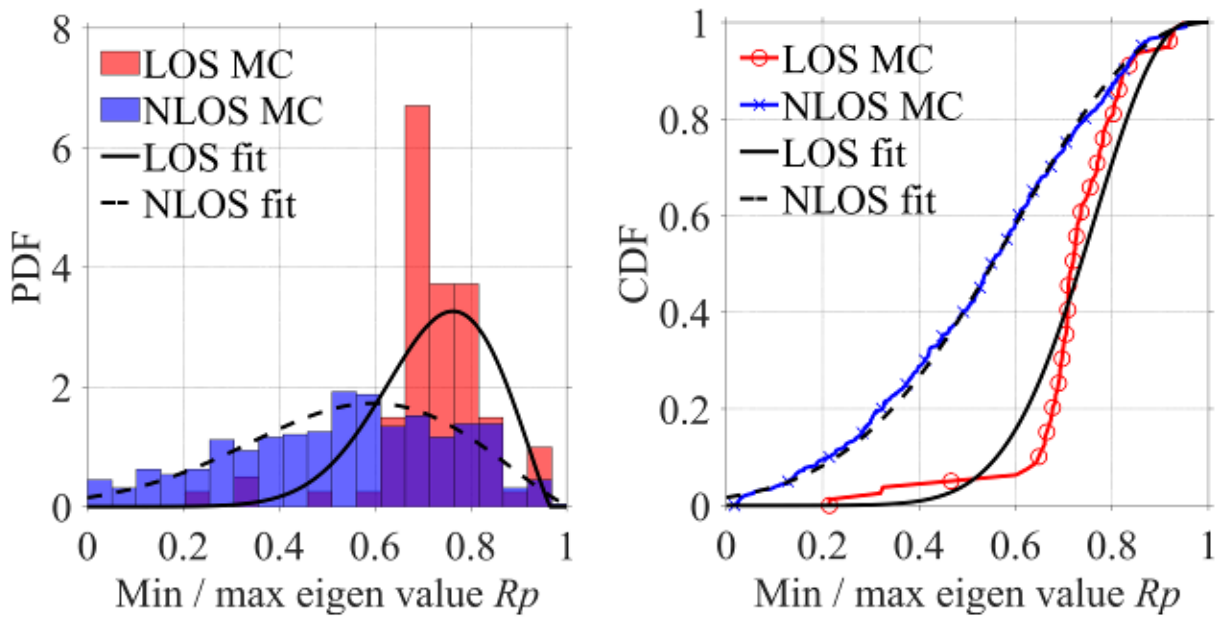

Fig. 14. PDF and CDF of measured covariance eigenvalue ratios $R_P$.

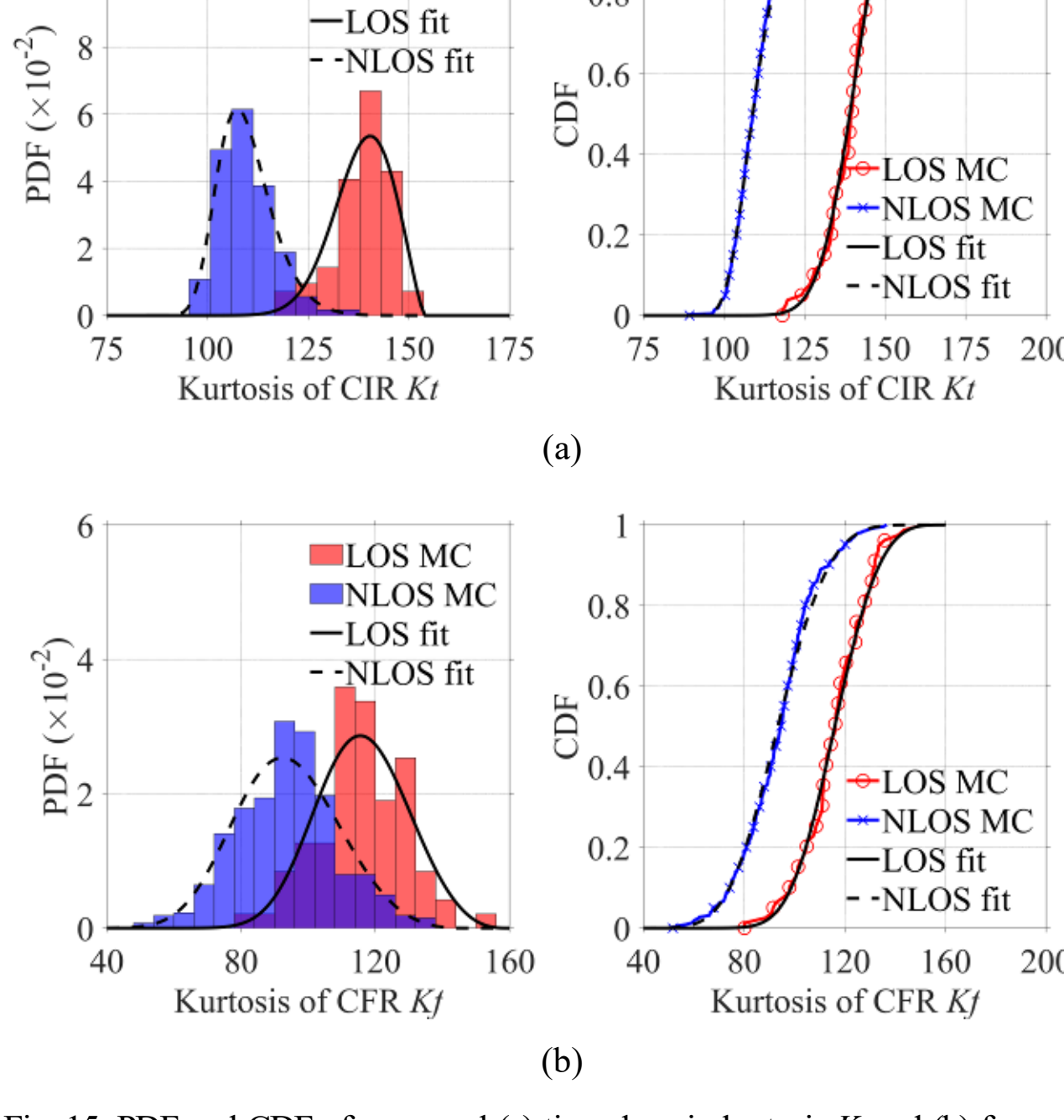

Fig. 15. PDF and CDF of measured (a) time-domain kurtosis $K_t$ and (b) frequency-domain kurtosis $K_f$.

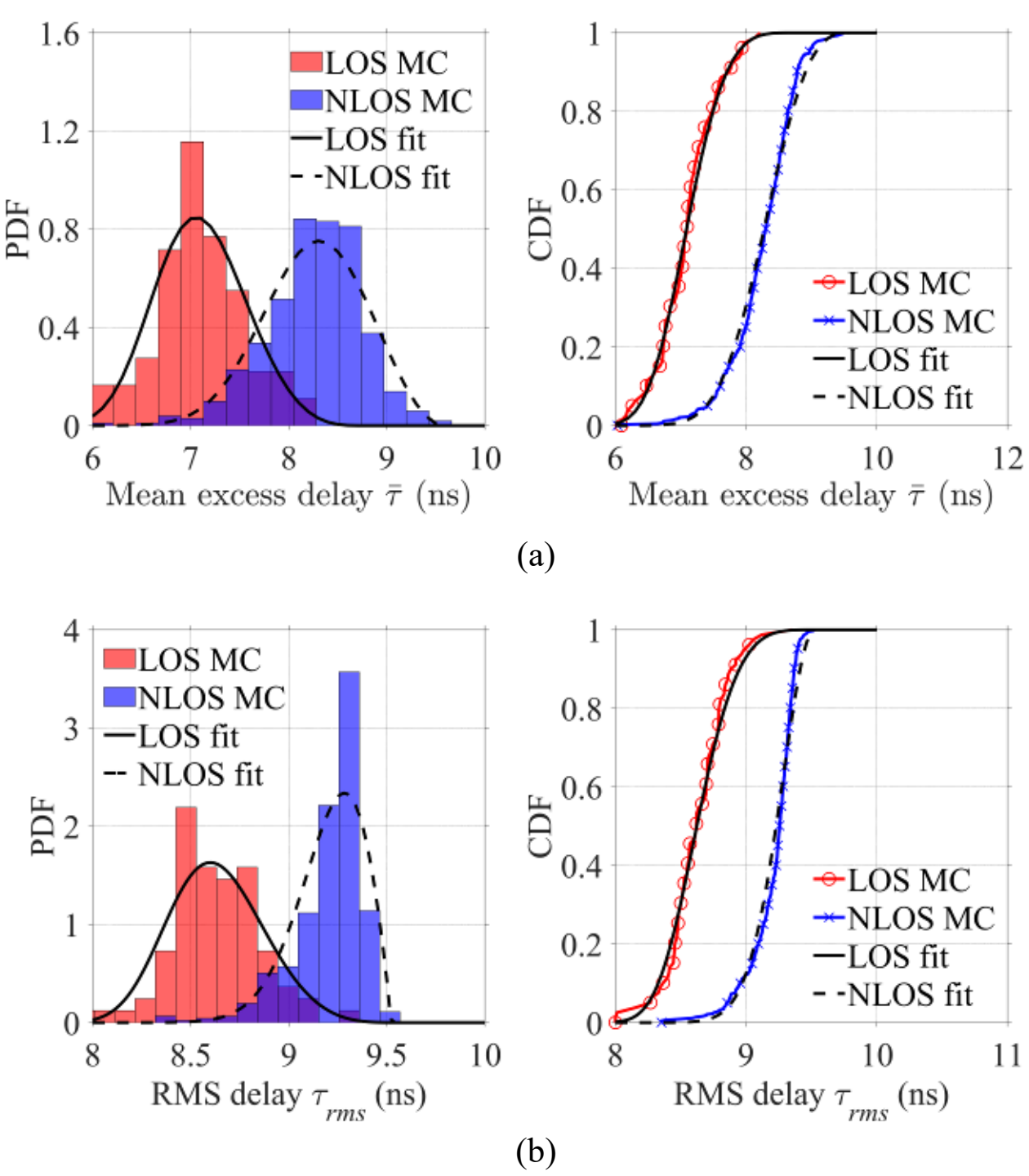

Fig. 16. PDF and CDF of measured (a) mean excess delay $\bar{\tau}$ and (b) RMS delay spread $\tau_{rms}$.

The PDF and CDF of the measured covariance eigenvalue ratio $R_P$ is shown in Fig. 14. The LOS eigenvalue ratio concentrates near 0.7[3] while the NLOS eigenvalue ratio is more spread in the 0-1 range. This is consistent with simulation results in Fig. 7 and a clear difference between LOS and NLOS behavior can be observed. The statistical distribution of the measured time-domain kurtosis $K_t$ and frequency-domain kurtosis $K_f$ are presented in Fig. 15. In Fig. 15 (a), $K_t$ statistics exhibit a similar behavior than in simulation (Fig. 8 (a)) 90% of $K_t$ values are smaller than 120 in NLOS clusters while it is only 4% in LOS clusters. The difference in $K_f$ distributions in LOS and NLOS clusters is lightly higher than in Fig. 8 (b). 65% of LOS propagation CFR kurtosis is smaller than 120.

[3] Unlike in simulations, the ratio differs from 1 as the antenna beamwidth is not identical in both angular planes.

The PDF and CDF of measured excess delay $\bar{\tau}$ and RMS delay spread $\tau_{rms}$ are shown in Fig. 16. Both measured average excessed delay and RMS delay spread of LOS propagation do not exhibit PDF as narrow as in simulations in Fig. 9. Indeed, in measurements, LOS clusters contain usually some multipath due to possible reflections in the vicinity of the antennas. However, while $\bar{\tau}$ behavior is similar than simulations, the difference between LOS and NLOS $\tau_{rms}$ distributions is more obvious in measurements than in simulations. Despite this, the shapes of the LOS and NLOS PDFs for all five metrics are different from each other and suggest a possible NLOS identification.

The parameters of the GEV distributions used for fitting and the RMSE for the five statistical features are shown in TABLE V. For all metrics, RMSE is lower than 0.08 in LOS and than 0.05 in NLOS.

TABLE V
MEASURED PARAMETERS OF GEV DISTRIBUTION

| Metrics | Parameters | LOS | NLOS |
|---|---|---|---|
| $R_P$ | $\gamma_P$ | -0.4704 | -0.4474 |
| | $\mu_P$ | 0.6918 | 0.4712 |
| | $\sigma_P$ | 0.1288 | 0.2400 |
| | RMSE | 0.0824 | 0.0103 |
| $K_t$ | $\gamma_t$ | -0.4421 | -0.0989 |
| | $\mu_t$ | 136.4 | 106.7 |
| | $\sigma_t$ | 7.722 | 5.995 |
| | RMSE | 0.0267 | 0.0073 |
| $K_f$ | $\gamma_f$ | -0.2298 | -0.2521 |
| | $\mu_f$ | 111.3 | 88.11 |
| | $\sigma_f$ | 13.42 | 15.02 |
| | RMSE | 0.0186 | 0.0178 |
| $\bar{\tau}$ | $\gamma_\tau$ | -0.2487 | -0.3547 |
| | $\mu_\tau$ | 6.931 | 8.093 |
| | $\sigma_\tau$ | 0.4484 | 0.5253 |
| | RMSE | 0.0241 | 0.0250 |
| $\tau_{rms}$ | $\gamma_{rms}$ | -0.2293 | -0.5289 |
| | $\mu_{rms}$ | 8.541 | 9.167 |
| | $\sigma_{rms}$ | 0.2322 | 0.1874 |
| | RMSE | 0.0253 | 0.0489 |

### *C. Performance of NLOS Identification*

TABLE VI
MEASURED PROBABILITIES OF ERROR TESTED FOR NLOS IDENTIFICATION

| Methods | Metrics | Type I | Type II |
|---|---|---|---|
| MLR | $R_P$ | 0.1050 | 0.2418 |
| | $K_t$ | 0.0200 | 0.0573 |
| | $K_f$ | 0.2598 | 0.1228 |
| | $\bar{\tau}$ | 0.1900 | 0.1111 |
| | $\tau_{rms}$ | 0.1750 | 0.1228 |
| | $R_P\ K_t\ K_f\ \bar{\tau}\ \tau_{rms}$ | 0.1319 | 0.0183 |
| ANN | $[R_P, K_t, K_f, \bar{\tau}, \tau_{rms}]$ | 0.0669 | 0.0408 |

To evaluate the performance of classification, the probabilities of type I and type II errors are shown in TABLE VI. The single metric that performs the best identification is $K_t$ like in simulation. More metrics benefit to reduce the probability of misidentification. Type I and II errors using MLR are 0.1319 and 0.0183 respectively. Type I and II errors with ANN are 0.07 and 0.04 respectively, which, unlike simulations, outperforms MLR for type I errors only.

Due to weaker differences between the LOS and NLOS behavior in measurements than in simulations, identification errors are slightly larger. Another possible source of degradation can be due to the horn antenna that exhibits some side-lobes in the E-plane, which can lead to interferences in the elevation plane of the measured PAS. Those interferences are not necessarily identified as clusters, such as the ones at $\varphi = 0°$ near the LOS cluster in Fig.13, but could anyway disturb the clusters' spatial behavior. Overall, performance appears reasonable. As a comparison, the CIR kurtosis, the mean excess delay, and the RMS delay spread used for identification at 28 GHz In [61] lead to total (type I+II) error probabilities of 0.3799, 0.2188, and 0.3360, respectively. Corresponding performance from TABLE VI are 0.0773, 0.3011, and 0.2978, respectively. Therefore, while kurtosis of an omnidirectional mm-wave channel in [59] led to poorer identification performance than time domain metrics, it is the opposite with the angular cluster-based identification proposed in this paper. This observation is also compliant with simulation results in section III-C. Results obtained with ANN are in the same order as the ones obtained experimentally in the 3.1-6.3 GHz in [31] where a 0.08 type I and a 0.09 type II errors are reported.

## V. CONCLUSION

Due to wideband and enhanced spatial properties (thanks to directive antennas), the millimeter band is an excellent candidate for accurate indoor localization. However, because of severe blockage effects at these frequencies, transmissions over indirect paths are necessary, thereby introducing large errors in the positioning process. To partly mitigate those errors, the nature of propagation, i.e., LOS or NLOS, over which the wireless link is established is a key indicator. This paper reports a method for LOS/NLOS identification of all individual clusters existing in the whole angular channel space of a 60 GHz communication using directive beam scanning antennas. Based on the beam training strategy in IEEE 802.11ad where both Tx and Rx antennas scan the whole angular space, a power angular spectrum (PAS) is obtained at the receiver. This readily available knowledge is here used to assess the spatial behavior of five different channel metrics, namely the angular covariance, the time-domain and frequency-domain channel kurtosis, the mean excess delay, and the RMS delay spread. Using these metrics, a noticeable difference between LOS and NLOS clusters is observed. The maximum likelihood ratio (MLR) and artificial neural network (ANN) are operated as classifiers. Training and testing are performed at 60 GHz, in simulation, using the IEEE 802.11ad conference room channel model and in measurement, using a VNA-based experimental setup. ANN is found to outperform MLR in most situations. Type I error, being a LOS transmission identified as NLOS, is, in simulation, about 0.05 using MLR and 0.02 using ANN, and in measurements, about 0.13 using MLR and 0.07 using ANN. Type II error, being an NLOS transmission identified as LOS, is, in simulation, about 0.03 with MLR and 0.003 with ANN, and in measurements, about 0.02 using MLR and 0.04 using ANN. These relatively low error values are similar to the ones obtained in the literature (e.g., [31], [32]) where identification is performed on quasi-omnidirectional channels, not considering the antenna's spatial filtering behavior in directional mm-wave transmission. Consequently, results obtained in this paper show that it is possible to perform NLOS identification of individual angular clusters at mm-wave frequencies in indoor environments. As a perspective to this work, further experiments should be conducted to assess the proposed approach in longer range scenarios. Furthermore, a deeper analysis on the optimal ANN architecture could be conducted depending on the considered scenario and different machine learning-based algorithm, such as random forest, should be compared in terms of identification performance

## References

[1] L. Atzori, A. Iera, and G. Morabito, "The internet of things: A survey," *Comput. Netw.,* vol. 54, no. 15, pp. 2787-2805, Oct. 2010.

[2] *3GPP release 15 description*, TR 21.915, Oct. 2019.

[3] *3GPP release 17 description*, TR 21.917, Jan. 2023.

[4] A. Maltsev, V. Erceg, and E. Perahia, "Channel models for 60 GHz wlan systems," IEEE, 802.11-09/0334r8, May. 2010.

[5] A. Maltsev, A. Pudeyev, and A. Gagiev, "Channel models for IEEE 802.11ay," IEEE, 802.11-15/1150r9, Mar. 2017.

[6] A. Maltsev, R. Maslennikov, A. Sevastyanov, A. Khoryaev, and A. Lomayev, "Experimental investigations of 60 GHz WLAN systems in office environment," *IEEE J. Sel. Areas Commun.,* vol. 27, no. 8, pp. 1488-1499, Oct. 2009.

[7] *ISO/IEC/IEEE international standard - information technology - telecommunications and information exchange between systems local and metropolitan area networks--specific requirements part 11: Wireless lan medium access control (MAC) and physical layer (PHY) specification*, IEEE Standard 802.11-2012, Nov. 2012.

[8] M. Jacob *et al.*, "Diffraction in mm and sub-mm wave indoor propagation channels," *IEEE Trans. Microw. Theory Tech.,* vol. 60, no. 3, pp. 833-844, Mar. 2012.

[9] T. Mavridis, L. Petrillo, J. Sarrazin, A. Benlarbi-Delai, and P. De Doncker, "Near-body shadowing analysis at 60 GHz," *IEEE Trans. Antennas Propag.,* vol. 63, no. 10, pp. 4505-4511, Oct. 2015.

[10] G. R. MacCartney, S. J. Deng, S. Sun, and T. S. Rappaport, "Millimeter-wave human blockage at 73 GHz with a simple double knife-edge diffraction model and extension for directional antennas," in *Proc. IEEE Veh. Technol. Conf.*, Montreal, QC, Canada, 18-21 Sep. 2016, pp. 1-6

[11] C. Slezak, V. Semkin, S. Andreev, Y. Koucheryavy, and S. Rangan, "Empirical effects of dynamic human-body blockage in 60 GHz communications," *IEEE Commun. Mag.,* vol. 56, no. 12, pp. 60-66, Dec. 2018.

[12] T. Y. Bai, R. Vaze, and R. W. Heath, "Analysis of blockage effects on urban cellular networks," *IEEE Trans. Wireless Commun.,* vol. 13, no. 9, pp. 5070-5083, Sep. 2014.

[13] M. K. Samimi, T. S. Rappaport, and G. R. MacCartney, "Probabilistic omnidirectional path loss models for millimeter-wave outdoor communications," *IEEE Wireless Commun. Lett.,* vol. 4, no. 4, pp. 357-360, Aug. 2015.

[14] S. Collonge, G. Zaharia, and G. El Zein, "Influence of the human activity on wide-band characteristics of the 60 GHz indoor radio channel," *IEEE Trans. Wireless Commun.,* vol. 3, no. 6, pp. 2396-2406, Nov. 2004.

[15] K. Venugopal and R. W. Heath, "Millimeter wave networked wearables in dense indoor environments," *IEEE Access,* vol. 4, pp. 1205-1221, 2016.

[16] Z. Sahinoglu, S. Gezici, and I. Güvenc, *Ultra-wideband positioning systems: Theoretical limits, ranging algorithms, and protocols*. The Edinburgh Building, Cambridge CB2 8RU, United Kingdom: Cambridge University Press, 2008, pp. 64-73.

[17] D. Dardari, A. Conti, U. Ferner, A. Giorgetti, and M. Z. Win, "Ranging with ultrawide bandwidth signals in multipath environments," *Proc. IEEE,* vol. 97, no. 2, pp. 404-426, Feb. 2009.

[18] C. K. Seow and S. Y. Tan, "Non-line-of-sight localization in multipath environments," *IEEE Trans. Mob. Comput.,* vol. 7, no. 5, pp. 647-660, May. 2008.

[19] A. Jafari *et al.*, "UWB interferometry TDOA estimation for 60-GHz OFDM communication systems," *IEEE Antennas Wirel. Propag. Lett.,* vol. 15, pp. 1438-1441, Dec. 2016.

[20] D. Dardari, P. Closas, and P. M. Djuric, "Indoor traScking: Theory, methods, and technologies," *IEEE Trans. Veh. Technol.,* vol. 64, no. 4, pp. 1263-1278, Apr. 2015.

[21] A. F. Molisch, "Ultra-wide-band propagation channels," *Proc. IEEE,* vol. 97, no. 2, pp. 353-371, Feb. 2009.

[22] R. C. Qiu, "A study of the ultra-wideband wireless propagation channel and optimum UWB receiver design," *IEEE J. Sel. Areas Commun.,* vol. 20, no. 9, pp. 1628-1637, Dec. 2002.

[23] A. F. Molisch *et al.*, "A comprehensive standardized model for ultrawideband propagation channels," *IEEE Trans. Antennas Propag.,* vol. 54, no. 11, pp. 3151-3166, Nov. 2006.

[24] J. Khodjaev, Y. Park, and A. S. Malik, "Survey of NLOS identification and error mitigation problems in UWB-based positioning algorithms for dense environments," *Ann Telecommun,* vol. 65, no. 5-6, pp. 301-311, Jun. 2010.

[25] Z. Xiao *et al.*, "Non-line-of-sight identification and mitigation using received signal strength," *IEEE Trans. Wireless Commun.,* vol. 14, no. 3, pp. 1689-1702, Nov. 2014.

[26] Z. Xiao *et al.*, "Identification and mitigation of non-line-of-sight conditions using received signal strength," in *Proc. IEEE Int. Conf. Wireless Mob. Comput., Netw. Commun.*, Lyon, France, 7-9 Oct. 2013, pp. 667-674

[27] M. Pätzold, *Mobile radio channels*, 2nd ed. The Atrium, Southern Gate, Chichester, West Sussex, PO19 8SQ, United Kingdom: John Wiley & Sons, Ltd, 2012.

[28] C. Tepedelenlioglu, A. Abdi, and G. B. Giannakis, "The ricean k factor: Estimation and performance analysis," *IEEE Trans. Wireless Commun.,* vol. 2, no. 4, pp. 799-810, Jul. 2003.

[29] L. J. Greenstein, D. G. Michelson, and V. Erceg, "Moment-method estimation of the ricean k-factor," *IEEE Commun. Lett,* vol. 3, no. 6, pp. 175-176, Jun. 1999.

[30] J. Borras, P. Hatrack, and N. B. Mandayam, "Decision theoretic framework for NLOS identification," in *Proc. IEEE Veh. Technol. Conf.*, Ottawa, Canada, 18-21 May. 1998, vol. 1-3, pp. 1583-1587

[31] S. Marano, W. M. Gifford, H. Wymeersch, and M. Z. Win, "NLOS identification and mitigation for localization based on UWB experimental data," *IEEE J. Sel. Areas Commun.,* vol. 28, no. 7, pp. 1026-1035, Sep. 2010.

[32] I. Guvenc, C. C. Chong, and F. Watanabe, "NLOS identification and mitigation for UWB localization systems," in *Proc. IEEE Wireless Commun. Netw. Conf.*, Hong Kong, China, 11-15 Mar. 2007, pp. 1573-1578

[33] C. Huang *et al.*, "Machine learning-enabled LOS/NLOS identification for MIMO systems in dynamic environments," *IEEE Trans. Wireless Commun.,* vol. 19, no. 6, pp. 3643-3657, Jun. 2020.

[34] T. V. Nguyen, Y. Jeong, H. Shin, and M. Z. Win, "Machine learning for wideband localization," *IEEE J. Sel. Areas Commun.,* vol. 33, no. 7, pp. 1357-1380, Jul. 2015.

[35] M. Heidari, N. A. Alsindi, and K. Pahlavan, "UDP identification and error mitigation in toa-based indoor localization systems using neural network architecture," *IEEE Trans. Wireless Commun.,* vol. 8, no. 7, pp. 3597-3607, Jul. 2009.

[36] F. Xiao, Z. Guo, H. Zhu, X. Xie, and R. Wang, "AmpN: Real-time LOS/NLOS identification with WiFi," in *Proc. IEEE Int. Conf. Commun.*, Paris, France, 21-25 May. 2017, pp. 1-7

[37] A. Decurninge *et al.*, "CSI-based outdoor localization for massive MIMO: Experiments with a learning approach," in *Proc. Int. Symp. Wireless Commun. Syst.*, Lisbon, Portugal, 28-31 Aug. 2018, pp. 1-6

[38] C. Huang *et al.*, "Artificial intelligence enabled radio propagation for communications—part ii: Scenario identification and channel modeling," *IEEE Trans. Antennas Propag.,* vol. 70, no. 6, pp. 3955-3969, Jun. 2022.

[39] C. Huang *et al.*, "Artificial intelligence enabled radio propagation for communications—part i: Channel characterization and antenna-channel optimization," *IEEE Trans. Antennas Propag.,* vol. 70, no. 6, pp. 3939-3954, Jun. 2022.

[40] I. Guvenc and C. C. Chong, "A survey on TOA based wireless localization and NLOS mitigation techniques," *IEEE Commun. Surv. Tutor.,* vol. 11, no. 3, pp. 107-124, 2009.

[41] M. R. Akdeniz *et al.*, "Millimeter wave channel modeling and cellular capacity evaluation," *IEEE J. Sel. Areas Commun.,* vol. 32, no. 6, pp. 1164-1179, Jun. 2014.

[42] M. Shafi *et al.*, "Microwave vs. Millimeter-wave propagation channels: Key differences and impact on 5g cellular systems," *IEEE Commun. Mag.,* vol. 56, no. 12, pp. 14-20, Dec. 2018.

[43] W. Roh *et al.*, "Millimeter-wave beamforming as an enabling technology for 5g cellular communications: Theoretical feasibility and prototype results," *IEEE Commun. Mag.,* vol. 52, no. 2, pp. 106-113, Feb. 2014.

[44] S. Sun, T. S. Rappaport, R. W. Heath, A. Nix, and S. Rangan, "MIMO for millimeter-wave wireless communications: Beamforming, spatial multiplexing, or both?," *IEEE Commun. Mag.,* vol. 52, no. 12, pp. 110-121, Dec. 2014.

[45] B. Q. Yang *et al.*, "Digital beamforming-based massive MIMO transceiver for 5g millimeter-wave communications," *IEEE Trans. Microw. Theory Tech.,* vol. 66, no. 7, pp. 3403-3418, Jul. 2018.

[46] X. Zhang, A. F. Molisch, and S.-Y. Kung, "Variable-phase-shift-based rf-baseband codesign for MIMO antenna selection," *IEEE Trans. Signal Process.,* vol. 53, no. 11, pp. 4091-4103, Nov. 2005.

[47] J. Y. Wang *et al.*, "Beam codebook based beamforming protocol for multi-gbps millimeter-wave wpan systems," *IEEE J. Sel. Areas Commun.,* vol. 27, no. 8, pp. 1390-1399, Oct. 2009.

[48] T. Nitsche *et al.*, "IEEE 802.11ad: Directional 60 GHz communication for multi-gigabit-per-second wi-fi," *IEEE Commun. Mag.,* vol. 52, no. 12, pp. 132-141, Dec. 2014.

[49] Y. Ghasempour, C. R. C. M. da Silva, C. Cordeiro, and E. W. Knightly, "IEEE 802.11ay: Next-generation 60 GHz communication for 100 Gb/s wi-fi," *IEEE Commun. Mag.,* vol. 55, no. 12, pp. 186-192, Dec. 2017.

[50] J. Kim and A. F. Molisch, "Fast millimeter-wave beam training with receive beamforming," *J Commun Netw-S Kor,* vol. 16, no. 5, pp. 512-522, Oct. 2014.

[51] D. De Donno, J. Palacios, and J. Widmer, "Millimeter-wave beam training acceleration through low-complexity hybrid transceivers," *IEEE Trans. Wireless Commun.,* vol. 16, no. 6, pp. 3646-3660, Jun. 2017.

[52] R. M. Buehrer, H. Wymeersch, and R. M. Vaghefi, "Collaborative sensor network localization: Algorithms and practical issues," *Proc. IEEE,* vol. 106, no. 6, pp. 1089-1114, Jun. 2018.

[53] P. Indirayanti, T. Ayhan, M. Verhelst, W. Dehaene, and P. Reynaert, "A mm-precise 60 GHz transmitter in 40 nm CMOS for discrete-carrier indoor localization," *IEEE J. Solid-St. Circ.,* vol. 50, no. 7, pp. 1604-1617, Jul. 2015.

[54] A. Jafari *et al.*, "TDOA estimation method using 60 GHz OFDM spectrum," *Int J Microw Wirel T,* vol. 7, no. 1, pp. 31-35, Feb. 2015.

[55] A. Shahmansoori, G. E. Garcia, G. Destino, G. Seco-Granados, and H. Wymeersch, "Position and orientation estimation through millimeter-wave MIMO in 5g systems," *IEEE Trans. Wireless Commun.,* vol. 17, no. 3, pp. 1822-1835, Mar. 2018.

[56] J. W. Zhao *et al.*, "Angle domain hybrid precoding and channel tracking for millimeter wave massive MIMO systems," *IEEE Trans. Wireless Commun.,* vol. 16, no. 10, pp. 6868-6880, Oct. 2017.

[57] H. Wymeersch, G. Seco-Granados, G. Destino, D. Dardari, and F. Tufvesson, "5g mm wave positioning for vehicular networks," *IEEE Wirel. Commun.,* vol. 24, no. 6, pp. 80-86, Dec. 2017.

[58] X. R. Cui, T. A. Gulliver, J. Li, and H. Zhang, "Vehicle positioning using 5g millimeter-wave systems," *IEEE Access,* vol. 4, pp. 6964-6973, Oct. 2016.

[59] X. L. Liang, Y. H. Jin, H. Zhang, and T. T. Lyu, "NLOS identification and machine learning methods for predicting the outcome of 60 GHz ranging system," *Chinese J Electron,* vol. 27, no. 1, pp. 175-182, Jan. 2018.

[60] X. L. Liang, H. Zhang, T. T. Lv, X. R. Cui, and T. A. Gulliver, "A novel scheme for NLOS identification using energy detector in 60 GHz systems," *Int J Futur Gener Co,* vol. 9, no. 6, pp. 151-164, Jun. 2016.

[61] A. Huang, L. Tian, T. Jiang, and J. H. Zhang, "NLOS identification for wideband mmwave systems at 28 GHz," in *Proc. IEEE Veh. Technol. Conf.*, Kuala Lumpur, Malaysia, 28 Apr. - 01 May. 2019, pp. 1-4

[62] B. Hu, H. Tian, and S. Fan, "Millimeter wave LOS/NLOS identification and localization via mean-shift clustering," in *Proc. IEEE Annu. Int. Symp. Pers., Indoor Mob. Radio Commun.*, Istanbul, Turkey, 8-11 Sep. 2019, pp. 1-6

[63] E. Kurniawan, L. Zhiwei, and S. Sun, "Machine learning-based channel classification and its application to IEEE 802.11ad communications," in *Proc. IEEE Global Commun. Conf.*, Singapore, 4-8 Dec. 2017,

[64] R. Charbonnier *et al.*, "Calibration of ray-tracing with diffuse scattering against 28-GHz directional urban channel measurements," *IEEE Trans. Veh. Technol.,* vol. 69, no. 12, pp. 14264-14276, Dec. 2020.

[65] J. Brady, N. Behdad, and A. M. Sayeed, "Beamspace MIMO for millimeter-wave communications: System architecture, modeling, analysis, and measurements," *IEEE Trans. Antennas Propag.,* vol. 61, no. 7, pp. 3814-3827, Jul. 2013.

[66] N. Czink *et al.*, "A framework for automatic clustering of parametric MIMO channel data including path powers," in *Proc. IEEE Veh. Technol. Conf.*, Montreal, Canada, 25-28 Sep. 2006, pp. 114-119

[67] P. Lyu, A. Benlarbi-Delaï, Z. Ren, and J. Sarrazin, "Angular clustering of millimeter-wave propagation channels with watershed transformation," *IEEE Trans. Antennas Propag.,* vol. 70, no. 2, pp. 1279-1290, Feb. 2022.

[68] C. Huang *et al.*, "A power-angle-spectrum based clustering and tracking algorithm for time-varying radio channels," *IEEE Trans. Veh. Technol.,* vol. 68, no. 1, pp. 291-305, Jan. 2019.

[69] M. Ramadan, V. Sark, J. Gutierrez, and E. Grass, "NLOS identification for indoor localization using random forest algorithm," in *Proc. Int. ITG Workshop Smart Antennas*, Bochum, Germany, 14-16 Mar. 2018, pp. 1-5

[70] J. C. Fan and A. S. Awan, "Non-line-of-sight identification based on unsupervised machine learning in ultra wideband systems," *IEEE Access,* vol. 7, pp. 32464-32471, 2019.

[71] C. H. Jiang *et al.*, "UWB NLOS/LOS classification using deep learning method," *IEEE Commun. Lett,* vol. 24, no. 10, pp. 2226-2230, Oct. 2020.

[72] B. Efron and R. J. Tibshirani, *An introduction to the bootstrap*. 29 West 35th Street, New York, NY 10001-2299, USA: Chapman and Hall/CRC, 1993.